\documentclass[twocolumn,twocolappendix]{aastex63}
\usepackage{amsmath}
\usepackage{makecell}
\usepackage{xcolor}
\definecolor{dgreen}{RGB}{26,148,49}
\definecolor{xlinkcolor}{cmyk}{1,1,0,0}
\hypersetup{linkcolor=xlinkcolor,citecolor=xlinkcolor,urlcolor=xlinkcolor}

\shorttitle{Probing Intra-Halo Light with Galaxy Stacking in CIBER Images}
\shortauthors{Cheng et al.}
\graphicspath{{./}{figures/}}

\begin{document}

\title{Probing Intra-Halo Light with Galaxy Stacking in CIBER Images}

\author[0000-0002-5437-0504]{Yun-Ting Cheng}
\address{California Institute of Technology, 1200 E. California Boulevard, Pasadena, CA 91125, USA}
\email{ycheng3@caltech.edu}

\author{Toshiaki Arai}
\address{Institute of Space and Astronautical Science, Japan Aerospace Exploration Agency, 
Kanagawa 252-5210, Japan}

\author[0000-0002-3886-3739]{Priyadarshini Bangale}
\address{Center for Detectors, School of Physics and Astronomy, Rochester Institute of Technology, 1 Lomb Memorial Drive, Rochester, New York 14623, USA}

\author{James J. Bock}
\address{California Institute of Technology, 1200 E. California Boulevard, Pasadena, CA 91125, USA}
\address{Jet Propulsion Laboratory, California Institute of Technology, 4800 Oak Grove Drive, Pasadena, CA 91109, USA}

\author[0000-0001-5929-4187]{Tzu-Ching Chang}
\address{Jet Propulsion Laboratory, California Institute of Technology, 4800 Oak Grove Drive, Pasadena, CA 91109, USA}
\address{California Institute of Technology, 1200 E. California Boulevard, Pasadena, CA 91125, USA}

\author{Asantha Cooray}
\address{Department of Physics and Astronomy, University of California, Irvine, CA 92697, U.S.A.}

\author[0000-0002-9330-8738]{Richard M. Feder}
\address{California Institute of Technology, 1200 E. California Boulevard, Pasadena, CA 91125, USA}

\author{Phillip M. Korngut}
\address{California Institute of Technology, 1200 E. California Boulevard, Pasadena, CA 91125, USA}

\author{Dae Hee Lee}
\address{Korea Astronomy and Space Science Institute (KASI), Daejeon 305-348, Republic of Korea}

\author{Lunjun Liu}
\address{California Institute of Technology, 1200 E. California Boulevard, Pasadena, CA 91125, USA}

\author[0000-0001-6066-5221]{Toshio Matsumoto}
\address{Institute of Space and Astronautical Science, Japan Aerospace Exploration Agency, 
Kanagawa 252-5210, Japan}

\author[0000-0002-5698-9634]{Shuji Matsuura}
\address{School of Science and Technology, Kwansei Gakuin University, Sanda, Hyogo 669-1337, Japan}

\author[0000-0001-9368-3186]{Chi H. Nguyen}
\address{Center for Detectors, School of Physics and Astronomy, Rochester Institute of Technology, 1 Lomb Memorial Drive, Rochester, New York 14623, USA}

\author[0000-0002-6468-8532]{Kei Sano}
\address{Kyushu Institute of Technology, 1-1 Sensui-cho, Tobata, Kitakyushu, Fukuoka 804-8550, Japan}

\author[0000-0001-7143-6520]{Kohji Tsumura}
\address{Department of Natural Science, Faculty of Science and Engineering, Tokyo City University, Tokyo, 158-8557, Japan}

\author[0000-0001-8253-1451]{Michael Zemcov}
\address{Center for Detectors, School of Physics and Astronomy, Rochester Institute of Technology, 1 Lomb Memorial Drive, Rochester, New York 14623, USA}
\address{Jet Propulsion Laboratory, California Institute of Technology, 4800 Oak Grove Drive, Pasadena, CA 91109, USA}

\begin{abstract}
We study the stellar halos of $0.2\lesssim z \lesssim 0.5$ galaxies with stellar masses spanning $M_*\sim 10^{10.5}$ to $10^{12}M_\odot$ (approximately $L_*$ galaxies at this redshift) using imaging data from the Cosmic Infrared Background Experiment (CIBER). A previous CIBER fluctuation analysis suggested that intra-halo light (IHL) contributes a significant portion of the near-infrared extragalactic background light (EBL), the integrated emission from all sources throughout cosmic history. In this work, we carry out a stacking analysis with a sample of $\sim$30,000 Sloan Digital Sky Survey (SDSS) photometric galaxies from CIBER images in two near-infrared bands (1.1 and 1.8 $\mu$m) to directly probe the IHL associated with these galaxies. We stack galaxies in five sub-samples split by brightness and detect an extended galaxy profile beyond the instrument point spread function (PSF) derived by stacking stars. We jointly fit a model for the inherent galaxy light profile plus large-scale one- and two-halo clustering to measure the extended galaxy IHL. We detect nonlinear one-halo clustering in the 1.8 $\mu$m band at a level consistent with numerical simulations.  By extrapolating the fraction of extended galaxy light we measure to all galaxy mass scales, we find $\sim$ 30\%/15\% of the total galaxy light budget from galaxies is at radius $r >$ 10/20 kpc, respectively. These results are new at near-infrared wavelengths at the $L_*$ mass scale and suggest that the IHL emission and one-halo clustering could have appreciable contributions to the amplitude of large-scale EBL background fluctuations.
\end{abstract}


\section{Introduction}\label{S:Introduction}
In the standard cosmological paradigm, galaxies grow hierarchically through merger and accretion. Galaxies accreting onto more massive systems become disrupted and stars stripped away from their parent galaxies become redistributed in the merged dark matter halo.  This results in extended stellar halos that are known to span tens or hundreds of kilo-parsecs. The stellar emission from this material is sometimes referred to as ``intra-halo light’’ (IHL), or in massive galaxy clusters as ``intra-cluster light’’ (ICL).

The properties of stellar halos across a wide range of mass scales have been extensively studied using analytical models \citep[e.g., ][]{2007ApJ...666...20P} and $N$-body simulations \citep[e.g., ][]{2005ApJ...635..931B,2007ApJ...668..826C,2009ApJ...699.1518R,2010MNRAS.406..744C,2013MNRAS.434.3348C,2015MNRAS.454.3185C,2016MNRAS.458.2371R,2018MNRAS.479.4004E}. Several observations have constrained the ICL content in galaxy clusters \citep[e.g., ][]{2004ApJ...617..879L,2015MNRAS.449.2353B,2005ApJ...618..195G,2007ApJ...666..147G,2005ApJ...618..195G} as well as stellar halos in lower mass systems by deeply imaging individual galaxies \citep[e.g., ][]{2009AJ....138.1417T,2010AJ....140..962M,2014PASP..126...55A,2014ApJ...782L..24V,2018MNRAS.475.3348H} or through stacking \citep[e.g., ][]{2005MNRAS.358..949Z,2014MNRAS.443.1433D,2019ApJ...874..165Z,2019MNRAS.487.1580W}.

An independent way to study the aggregate emission from diffuse sources like IHL is through measurements of the extragalactic background light (EBL), which encodes the integrated emission from all sources across cosmic history \citep{2016RSOS....350555C}.  Absolute optical and near-infrared EBL photometry has proven challenging as measurements must tightly control systematic errors and carefully model and subtract local foregrounds \citep[e.g., ][]{2017PASJ...69...31K,2017NatCo...815003Z,2017ApJ...839....7M,2018AJ....156...86M,2021ApJ...906...77L}. Several authors  \citep{2007ApJ...666..663B, 2007ApJ...666...34L,2013PASJ...65..121T,2015ApJ...807...57M,2015ApJ...811...77S,2017NatCo...815003Z,2017ApJ...839....7M,2020ApJ...901..112S,2021ApJ...906...77L} have reported potential detections 
above the integrated galaxy light (IGL) derived from galaxy counts \citep{2010ApJ...723...40K,2011MNRAS.410.2556D,2012ApJ...752..113H,2016ApJ...827..108D,2020arXiv201203035S,2021MNRAS.503.2033K}, which may indicate the existence of extragalactic emission missed in source-counting surveys.

Additionally, EBL fluctuation analyses have also consistently reported excess fluctuations over those expected from the IGL \citep{2005Natur.438...45K,2007ApJ...666..658T,2011ApJ...742..124M,2012ApJ...753...63K,2012Natur.490..514C,2014Sci...346..732Z,2015NatCo...6.7945M,2015ApJ...807..140S,2019PASJ...71...82K,2019PASJ...71...88M}. One explanation is emission from the epoch of reionization \citep{2005Natur.438...45K,2011ApJ...742..124M,2012ApJ...753...63K,2015NatCo...6.7945M}, while other studies suggest IHL contributes most of the excess fluctuations \citep{2012Natur.490..514C}. In particular,  \cite{2014Sci...346..732Z} interpret imaging data from the Cosmic Infrared Background Experiment (CIBER) as arising from an IHL intensity comparable to the IGL at near-infrared wavelengths.  This result would imply that stars diffusely scattered in dark matter halos may account for a non-negligible fraction of the near-IR cosmic radiation budget. The absorption spectra from blazars constrain the EBL column density along the line of sight \citep[e.g., ][]{2006Natur.440.1018A,2007A&A...475L...9A,2008Sci...320.1752M,2010ApJ...723.1082A,2012Sci...338.1190A,2017A&A...606A..59H,2018ApJS..237...32A,2019ApJ...885..150A,2019MNRAS.486.4233A,2020MNRAS.494.5590A}. While IHL is generally produced at low redshifts, improving the uncertainties in its redshift history helps place IHL in the context of these constraints.

In this work, we further constrain the IHL using CIBER broadband imaging. Rather than studying EBL intensity fluctuations as in \citet{2014Sci...346..732Z}, we perform a stacking analysis to directly probe the stellar halos around galaxies. We stack a sample of $\sim 30,000$ Sloan Digital Sky Survey (SDSS) photometric galaxies at $z\sim$ 0.2-0.5 across five $2\times 2$ deg$^2$ fields. Our samples span a range of stellar masses at approximately $L_*$ scales at this redshift \citep{2013ApJ...777...18M}. Although we only study stellar halos around a subset of galaxies rather than the aggregate population as probed by fluctuations, stacking provides a direct path to probe the IHL associated with this sample. Stacking complements fluctuation measurements by probing the relationship between individual galaxies and their stellar halos. Stacking also allows us to investigate how stellar halos depend on host-galaxy properties, e.g, stellar mass, redshift, etc.  A complementary fluctuation analysis of these same data is currently in progress.

This paper is organized as follows. First, we introduce CIBER in Sec.~\ref{S:CIBER} and the data processing in Sec.~\ref{S:Reduction}. Sec.~\ref{S:Catalogs} and Sec.~\ref{S:MICECAT} describe the external data sets used in this work, including observed and simulated source catalogs. Sec.~\ref{S:stacking} details the stacking procedure and Sec.~\ref{S:PSF} describes the point spread function (PSF) model. The stacking results are presented in Sec.~\ref{S:Galaxy_Stacking}. Sec.~\ref{S:Model_Signal} introduces the theoretical model we use to fit the data and the parameter fitting procedure. The results on model parameter constraints are given in Sec.~\ref{S:Results}, and further discussion is presented in Sec.~\ref{S:Discussion}. Sec.~\ref{S:Conclusion} summarizes the paper. Throughout this work, we assume a flat $\Lambda$CDM cosmology with $n_s=0.97$, $\sigma_8=0.82$, $\Omega_m=0.26$, $\Omega_b=0.049$, $\Omega_\Lambda=0.69$, and $h=0.68$, consistent with the measurement from Planck \citep{2016A&A...594A..13P}. All fluxes are quoted in the AB magnitude system.

\begin{deluxetable*}{c|ccccccc}[ht!]
\tablenum{1}
\label{T:fields}
\tablecaption{CIBER Observing Fields}
\tablewidth{0pt}
\tablehead{
\colhead{Field Name} & \colhead{R.A. ($^{\circ}$)} & \colhead{Decl. ($^{\circ}$)} & \colhead{Time After Launch (s)} &  \colhead{Number of Frames Used} & \colhead{Integration Time (s)}
}
\startdata
Elat10 & 191.50 & 8.25 & 387-436 & 24 & 42.72 \\
Elat30 & 193.94 & 28.00 & 450-500 & 9 & 16.02\\
BootesB & 218.11 & 33.18 & 513-569 & 29 & 51.62 \\
BootesA & 219.25 & 34.83 & 581-636 & 28 & 49.84 \\
SWIRE(ELAIS-N1) & 241.53 & 54.77 & 655-705 & 25 & 44.50\\
\enddata
\tablecomments{We discard the beginning part of the Elat30 field integration due to pointing instability.}
\end{deluxetable*}

\section{CIBER Experiment}\label{S:CIBER}
CIBER\footnote{\url{https://ciberrocket.github.io/}} \citep{2013ApJS..207...31Z} is a rocket-borne instrument designed to characterize the near-infrared EBL. CIBER consists of four instruments: two wide-field imagers \citep{2013ApJS..207...32B}, a narrow-band spectrometer \citep{2013ApJS..207...34K}, and a low-resolution spectrometer \citep{2013ApJS..207...33T}. CIBER has flown four times in 2009 February, 2010 July, 2012 March, and 2013 June. The first three CIBER flights were launched at White Sands Missile Range, New Mexico on a Terrier-Black  Brant IX rocket. These flights reached $\sim$ 330 km apogee with $\sim$ 240 s of exposure time, and the payload was recovered for future flights. The fourth flight was a non-recovery flight launched 3:05 UTC 2013 June 6 from Wallops Flight Facility, Virginia on a four-stage Black Brant XII rocket. The payload reached 550 km altitude, much higher than the two-stage rocket used in the previous three flights. This gives more exposure time (335 s) for observing more science fields with long integrations to achieve better sensitivity and systematics control.

This work presents the first science results from the CIBER fourth flight imager data. The data from previous flights have been studied with a fluctuation analysis, published in \cite{2014Sci...346..732Z}. With a large field of view and low sky background above the atmosphere, CIBER imaging provides fidelity on angular scales from $7''$ to $2^{\circ}$. For stacking, CIBER imaging can trace low-surface-brightness emission on degree angular scales providing a unique dataset compared with ground-based or small field-of-view space-borne studies. Each CIBER imager uses a $1024\times 1024$ pixel HAWAII-1 HgCdTe detector. The two imagers are identical except for their $\lambda/\Delta \lambda\sim2$ filters, which are centered at $1.05$ and $1.79$ $\mu$m\footnote{In the first and second CIBER flights, the longer wavelength band is centered at 1.56 $\mu$m, and thus it is named the 1.6 $\mu$m band in previous CIBER publications \citep{2013ApJS..207...32B,2014Sci...346..732Z}.}.

During its fourth flight, CIBER observed 8 science fields with $\sim 50$ s integrations sampled at 1.78 s intervals. We discard the first three fields in this analysis due to contamination from airglow that produces a strong non-uniform emission across the images that requires aggressive filtering which also significantly reduces our signal \citep{2014Sci...346..732Z}. Table \ref{T:fields} summarizes the sky coordinates and the integration time of the five science fields used in this work. In the beginning of the Elat30 integration, the rocket's pointing was not stable which has the effect of smearing the PSF on the sky. As a result, we only use the last 16 s of this integration in our analysis.

\section{Data Processing}\label{S:Reduction}
\begin{figure*}[ht!]
\includegraphics[width=\linewidth]{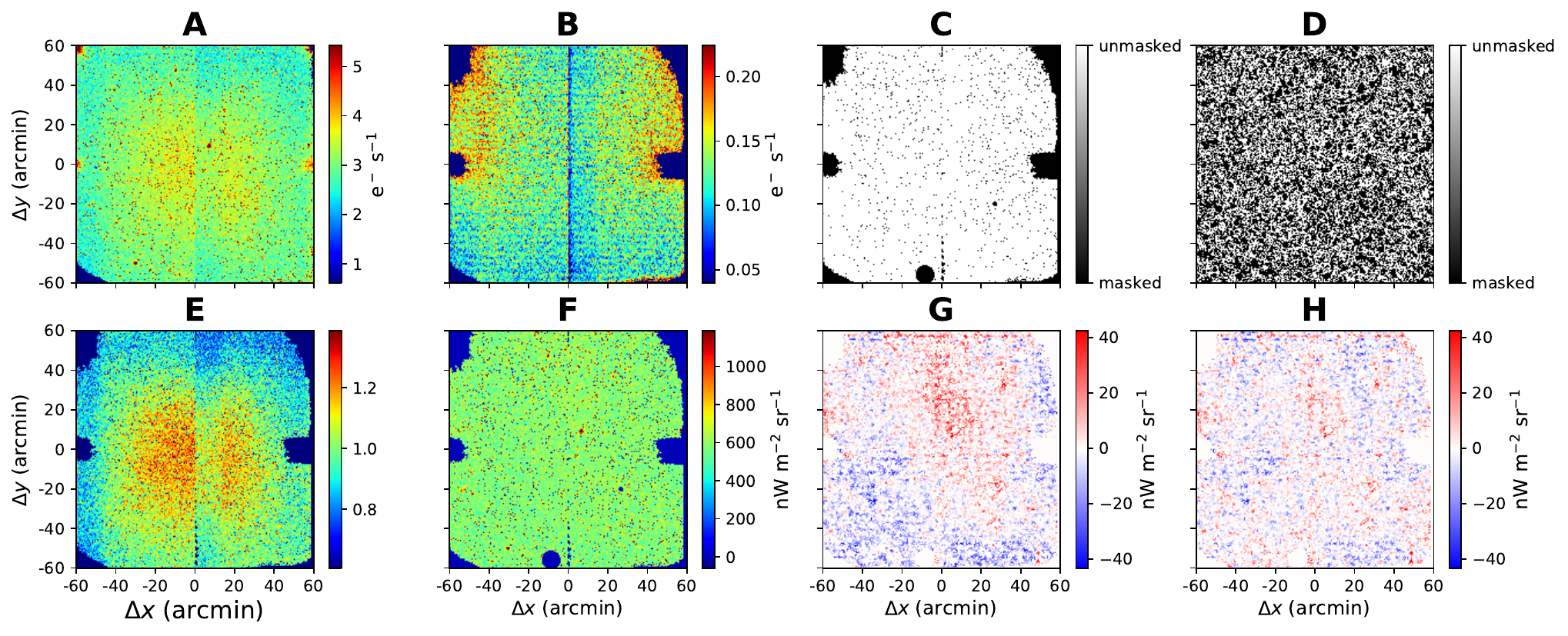}
\caption{\label{F:maps1} Images from the SWIRE field in the 1.1 $\mu$m band. A: the raw image of the photoccurent map. B: dark-current template constructed from dark images before the flight. C: instrument mask encoding the pixels with fabrication defects, unusual photocurrents, and cosmic-ray contamination. D: source mask for bright stars and galaxies in the 2MASS and Pan-STARRS catalogs. E: flat-field estimator from averaging the other four sky fields. F: raw image after dark-current subtraction, flat-field correction, and calibration. G: image in Panel F after (constant) background removal and masking. This image is smoothed with a $\sigma=35^{''}$ Gaussian kernel to highlight large-scale fluctuations. H: image in Panel G after subtracting a fitted 2D polynomial, also shown smoothed with a $\sigma=35^{''}$ Gaussian kernel. Compared to Panel G, we see that the large-scale background fluctuations have been reduced after filtering. This is the final product of the data reduction pipeline.}
\includegraphics[width=\linewidth]{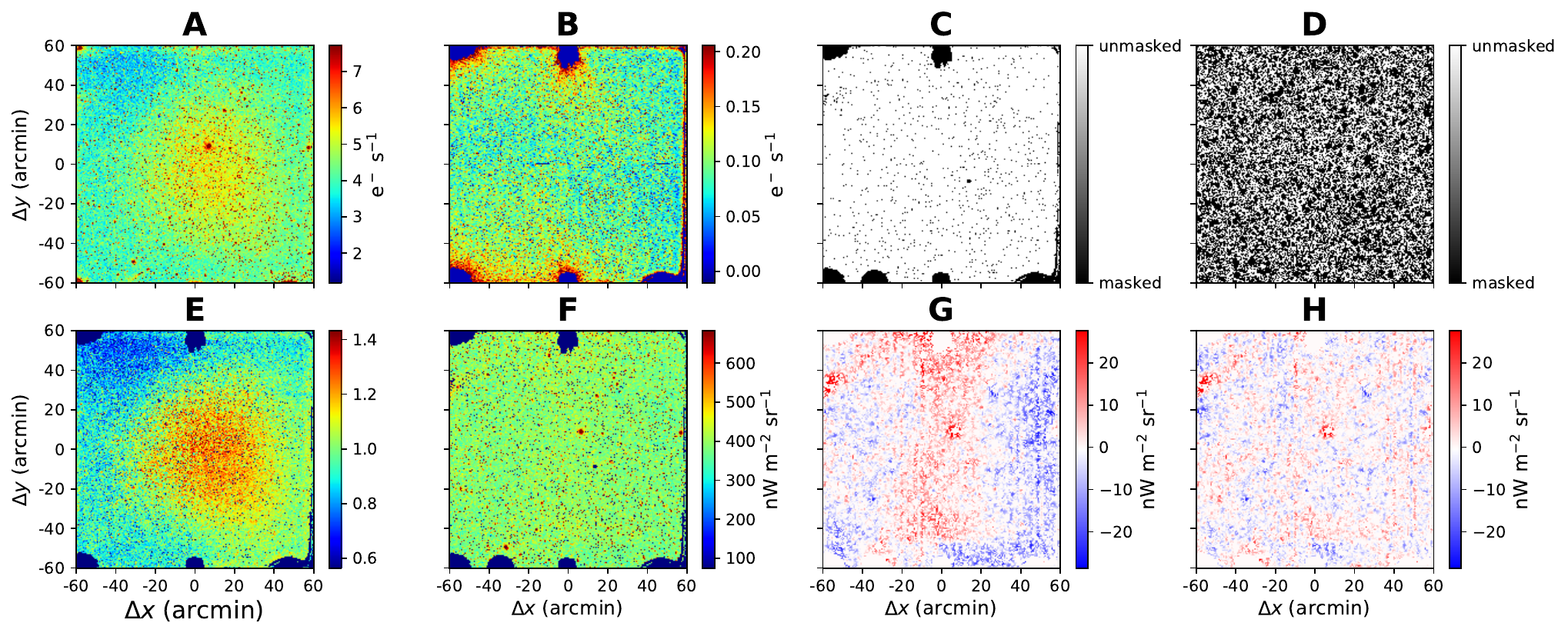}
\caption{\label{F:maps2} Same as Fig.~\ref{F:maps1} in the CIBER 1.8 $\mu$m band.}
\end{figure*}

In this section, we describe the data reduction from the raw flight data to the final images used for stacking.

\subsection{Raw Time Stream to Images}
The raw imager data provides a time series for each pixel. We fit a slope to the time stream to obtain the photocurrent in each pixel and convert the values from the raw analog-to-digital units (ADU) to e$^-$ s$^{-1}$ using known array gain factors.

The HAWAII-1 detector is linearly responsive to incoming flux over a certain dynamic range. For pixels pointing at bright sources, the detectors saturate and have a nonlinear flux dependence even for short integrations \citep{2013ApJS..207...32B}. In any pixel that collects more than 5000 ADU over the full integration only the first four frames are used in the photocurrent estimate. Hereafter, the term ``raw image'' refers to the photocurrent map after this linearity correction. Panel A of Fig.~\ref{F:maps1} and Fig.~\ref{F:maps2} show the raw images of the SWIRE field in the CIBER 1.1 and 1.8 $\mu$m bands, respectively. 

\subsection{Dark Current}\label{S:Dark_Current}
In the absence of incoming photons, the detectors have a nonzero response, commonly referred to as ``dark current,'' due to thermally produced charge carriers and multiplexer glow. The detector dark current is measured before each flight with the telescopes' cold shutters closed. We obtain a dark-current template for each detector by averaging 11 dark images and then subtracting each template from the corresponding raw images. The dark-current level in CIBER imagers is $\sim 0.1$ e$^-$ s$^{-1}$, less than 10 \% of the sky brightness. Panel B of Fig.~\ref{F:maps1} and Fig.~\ref{F:maps2} show the dark-current maps of CIBER 1.1 and 1.8 $\mu$m bands, respectively.

\subsection{Pixel Masks}\label{S:Masking}
We mask pixels that meet at least one of the following conditions: (1) a fabrication defect, (2) poor time-stream behavior, (3) abnormal photocurrents compared with other pixels, (4) a cosmic ray strike, or (5) being on or close to bright point sources on the sky. The pixels satisfying criteria (1)--(4) comprise the ``instrument mask'' and a ``source mask'' is composed of pixels with condition (5).

\subsubsection{Instrument Mask}\label{S:Instrument_Mask}
Pixels with fabrication defects and significant multiplexer glow are mostly distributed near the edges or corners of each quadrant on the detector arrays. They exhibit pathologies in their photocurrent response and can be found by comparison to the population of normal pixels.  We perform a 3$\sigma$ clipping on stacked dark images (the same dataset used for a dark-current template in Sec.~\ref{S:Dark_Current}) to identify these pixels. 

During integration, some cosmic-ray events or electronic transients leave a step feature in the time stream. We use a 100$\sigma$ clip on each time stream to pick out pixels that show these abrupt changes during an integration. Sometimes cosmic-ray events also leave a comet-like structure on the array and these regions are also masked. The union of the pathological pixel, time-stream masks, and cosmic-ray masks form the instrument mask. In total, $\sim 10 \%$ of pixels are removed by the instrument mask. Panel C of Fig.~\ref{F:maps1} and Fig.~\ref{F:maps2} show the instrument masks in the SWIRE field of the 1.1 and 1.8 $\mu$m bands, respectively.

\subsubsection{Source Mask}\label{S:Source_Mask}
To remove bright foreground stars and galaxies in our fields, we use position and brightness information from the Pan-STARRS and 2MASS catalogs (see Sec.~\ref{S:Catalogs} for details). We further derive source magnitudes in the two CIBER bands, $m_{\rm 1.1}$ and $m_{\rm 1.8}$, from these catalogs, as detailed in Sec.~\ref{S:Catalogs}. We mask all point sources brighter than $m_{\rm 1.1}=20$, choosing a masking radius for each source derived as follows. With the modeled instrument PSF (Sec.~\ref{S:PSF_instr}), the masking radius is chosen such that for each source, pixels with intensity brighter than $\nu I_\nu^{\rm th} = 1$ nW m$^{-2}$ sr$^{-1}$ in the 1.1 $\mu$m band are masked. This choice of threshold value removes $\sim 50 \%$ of pixels in each field. We apply the same masking radius to 1.8 $\mu$m band sources. The same masking function is also applied to simulations to account for residual emission from bright sources outside the masks and the unmasked faint populations. Panel D of Fig.~\ref{F:maps1} and Fig.~\ref{F:maps2} shows the SWIRE field source mask in the CIBER 1.1 and 1.8 $\mu$m bands, respectively.

The final mask we apply to the data is the union of the instrument mask and source mask.  After applying these masks, we apply a final 3$\sigma$ pixel clipping mask to identify additional outliers not flagged through the other methods (e.g., from low-energy cosmic-ray events or electronic tranisents).

\subsection{Flat-Fielding}\label{S:FF}
CIBER images have a non-uniform response to a constant sky brightness across the detector array, known as the flat-field response. For each CIBER field, the flat-field is estimated by averaging the dark-current-subtracted flight images of the other four sky fields.

A laboratory flat-field measurement was also taken before the flight using a field-filling integrating sphere, a uniform radiance source with a solar spectrum \citep[described in][]{2013ApJS..207...32B}. Ideally, this is a better approach to measure the flat-field since the one derived from stacking flight images contains fluctuations from the other fields that will not average down completely due to the small number of images. However, we found the flat-field from the integrating sphere is not consistent with the flight data on large spatial scales \citep[see ][]{2014Sci...346..732Z}, and therefore we do not use it in our analysis. The flat-field estimator for the SWIRE field in the CIBER 1.1 and 1.8 $\mu$m bands is shown in Panel E of Fig.~\ref{F:maps1} and Fig.~\ref{F:maps2}, respectively.

\subsection{Surface Brightness Calibration}\label{S:Calibration}
Throughout this work, we use nW m$^{-2}$ sr$^{-1}$ for the units of surface brightness ($\nu I_{\nu}$). The calibration factor, $C$, which converts photocurrent (e$^-$ s$^{-1}$) to intensity (nW m$^{-2}$ sr$^{-1}$) is derived in the following steps: 
\begin{enumerate}
    \item Take the raw images, subtract the dark-current template, correct for the flat-field, and apply the instrument and source masks;
    \item Subtract the mean photocurrent in the unmasked region;
    \item For each star in the Pan-STARRS catalog, calculate the flux $\nu F_\nu$ in CIBER bands from $m_{\rm 1.1}$ and $m_{\rm 1.8}$;
    \item Sum the photocurrent in a 5$\times$5 stamp centered on the source position\footnote{We have tested that using 3$\times$3, 5$\times$5, or 7$\times$7 stamp size gives consistent results. Our beam size is approximately twice of the pixel size, so a 3$\times$3 stamp already has enclosed most of the flux from a point source.};
    \item Repeat steps (3) and (4) for all the selected stars (see below) and take the average value of the flux ratio from (3) and (4) as the calibration factor $C$.
\end{enumerate}

We select stars in the magnitude range $12.5 < m_{\rm 1.1} < 16$ for the 1.1 $\mu$m band, and $13.5 < m_{\rm 1.1} < 17$ for the 1.8 $\mu$m band. These magnitude ranges are chosen such that the brightest sources that saturate the detectors (even after nonlinear correction) are excluded. Faint sources are not used because of their low signal-to-noise ratio. We use a different magnitude range for each band as they have different point-source sensitivities. Panel F of Fig.~\ref{F:maps1} and Fig.~\ref{F:maps2} shows the SWIRE field images masked by instrument masks at 1.1 and 1.8 $\mu$m, respectively, after flat-fielding and calibration.

\subsection{Background Removal}
The total sky emission is composed of the EBL and various foreground components, including zodiacal light (ZL), diffuse galactic light (DGL), and integrated starlight (ISL) from the Milky Way \citep{2014Sci...346..732Z,2017ApJ...839....7M}. ZL is the dominant foreground, approximately an order of magnitude brighter than the EBL \citep{2017ApJ...839....7M}. Nevertheless, with its smooth spatial distribution on degree scales, the ZL can be mostly removed by subtracting the mean sky brightness in each field. Panel G of Fig.~\ref{F:maps1} and Fig.~\ref{F:maps2} shows the mean-subtracted and masked SWIRE images at 1.1 and 1.8 $\mu$m, respectively. To highlight the large-scale fluctuations, we smooth the images with a $\sigma=35^{''}$ Gaussian kernel.

\subsection{Image Filtering}\label{S:Image_Filtering}
Although the ZL signal is smooth, a flat-field estimation error may induce a non-uniform ZL residual that cannot be removed by mean subtraction. This residual may dominate over cosmological fluctuations on large scales. Therefore, after removing the mean value in the image, we filter the images by fitting and subtracting a third/fifth-order 2D polynomial function for the 1.1/1.8 $\mu$m images to filter out any residual large-scale variations (Panel H of Fig.~\ref{F:maps1} and Fig.~\ref{F:maps2}). The filtering will also suppress large-scale cosmological signals and therefore the choice of polynomial order used for filtering is determined by optimizing the trade-off between the reduction of background fluctuations and the large-scale two-halo signal. The effect of filtering on the detected one-halo and galaxy extension terms is small, as our filtering removes fluctuations at a much larger scales than these signals and the signal filtering is accounted for in simulations (see Sec.~\ref{S:Model_Signal}).

\section{External Catalogs}\label{S:Catalogs}
Throughout this work, we used several external source catalogs for (1) masking bright foreground sources (Sec.~\ref{S:Source_Mask}), (2) calibration (Sec.~\ref{S:Calibration}), (3) modeling the PSF by stacking bright stars in the fields (Sec.~\ref{S:PSF}), and (4) selecting galaxies for stacking (Sec.~\ref{S:Galaxy_Stacking}). 

To match the catalog sources to our data, we fit the astrometry coordinates of our images with the online software \url{nova.astrometry.net} \citep{2010AJ....139.1782L}. For each image, we solve for the astrometry in four quadrants separately to mitigate the effect of image distortion. Since there is a fixed $\sim 50^{''}$ misalignment between the 1.1 and 1.8 $\mu$m images as they are produced by different telescopes, their astrometry is solved separately.

\subsection{Pan-STARRS}
We use the Pan-STARRS catalog \citep{2016arXiv161205560C} for masking. Pan-STARRS covers all of the CIBER fields with a depth of $m\sim 20$ in the \textit{g, r, i, z, y} bands. We query the source positions and magnitudes in all five Pan-STARRS bands from their DR1 {\tt\string MeanObject} table and derive $m_{\rm 1.1}$ and $m_{\rm 1.8}$ with the \textit{LePhare} SED-fitting software \citep{1999MNRAS.310..540A,2006A&A...457..841I}. We use sources that have a y-band measurement and a quality flag ({\tt\string qualityFlag} in {\tt\string ObjectThin} table) that equals to 8 or 16 for masking.

\subsection{2MASS}
Some bright stars are not included in the Pan-STARRS catalog, and thus we use the 2MASS \citep{2006AJ....131.1163S} {\tt\string Point Source Catalog (PSC)} to get the complete point-source list. For 2MASS sources, $m_{\rm 1.1}$ ($m_{\rm 1.8}$) is derived by linear extrapolation with the 2MASS photometric fluxes in the \textit{J} and \textit{H} (\textit{H} and \textit{K$_s$}) bands, respectively. We also use bright stars in 2MASS for modeling the PSF (see Sec.~\ref{S:PSF}).

\subsection{SDSS}
We use the Sloan Digital Sky Survey (SDSS) DR13 \citep{2017AJ....154...28B} {\tt\string PhotoObj} catalog to get the star/galaxy classification (``type'' attribute 6--stars, 3--galaxies) and the galaxy photometric redshift (``Photoz'' attribute) for sources in our fields. This information is essential for selecting target galaxies for stacking and inferring their redshift distribution (Sec.~\ref{S:Source_Selection}), as well as selecting stars for stacking to model the PSF (Sec.~\ref{S:PSF}).

\subsection{SWIRE Photometric Redshift Catalog}
\citet{2008MNRAS.386..697R,2013MNRAS.428.1958R} performed SED fitting on $\sim 10^6$ sources in the SWIRE field, based on optical and infrared photometric data from multiple surveys. This provides information on the stellar masses of our stacked galaxies for our analysis (see Sec.~\ref{S:Source_subsamples}).

\subsection{Gaia}
Gaia DR2 \citep{2016A&A...595A...1G,2018A&A...616A...1G} provides high-precision astrometry for stars in the Milky Way, which gives high-purity star samples used for both validating the PSF model (Sec.~\ref{S:PSF_validation}) and cleaning residual stars in the galaxy sample selected by SDSS (Sec.~\ref{S:Source_Selection}).

\subsection{Nearby Cluster Catalog}\label{S:Nearby_Clusters}
Nearby galaxy clusters along the line of sight introduce extended emission in stacking, so we exclude galaxies that are close to nearby clusters (Sec.~\ref{S:Source_Selection}). We use the cluster catalog from \citet{2012ApJS..199...34W}, which compiles $0.05 \leqslant z < 0.8$ galaxy clusters detected in SDSS-III \citep{2011ApJS..193...29A}. We also use the Abell cluster samples \citep{1958ApJS....3..211A} for local galaxy clusters. There are 7 Abell clusters and $\sim$ 200 clusters from \citet{2012ApJS..199...34W} over the five CIBER fields.

\section{Simulation Catalog---MICECAT}\label{S:MICECAT}
In addition to the observed source catalogs, we make use of the MICECAT simulated galaxy catalog \citep{2015MNRAS.448.2987F,2015MNRAS.447.1319F,2015MNRAS.447.1724H} to estimate the signal from galaxy clustering. MICECAT is a product of the $N$-body cosmological simulation MICE Grand Challenge run (MICE-GC), which has 70 billion dark matter particles in a 3072$^3$ Mpc$^3$h$^{-3}$ cubic co-moving box. The dark matter halos are resolved down to $\sim 3 \times 10^{10} M_\odot h^{-1}$.

MICECAT is a mock catalog that simulates ideal observations of a 5000 deg$^2$ light cone covering $0<z<1.4$. MICECAT builds on MICE-GC by combining a halo occupation distribution (HOD) with subhalo abundance matching (SHAM) to calibrate to observed luminosity functions and clustering \citep{2015MNRAS.447..646C}. MICECAT simulates a mass-limited sample complete to $m_i\sim 22$ and $m_i\sim 24$ at $z\simeq 0.5$ and $z\simeq 0.9$, respectively \citep{2015MNRAS.453.1513C}.  The MICECAT mocks are large enough to permit us to generate up to $\sim10^3$ independent CIBER field-sized ($2\times 2$ deg$^2$) mock catalogs. We use modeled magnitudes from MICECAT in Euclid NISP \textit{Y} and \textit{H} bands for CIBER $m_{\rm 1.1}$ and $m_{\rm 1.8}$, respectively, since the NISP filters are similar to the CIBER imager bands.

MICECAT simulates both central and satellite galaxies generated with its HOD+SHAM model, which allows us to model the linear (two-halo) and nonlinear (one-halo) clustering in the stacking signal separately.  We use the radial shapes derived from MICECAT stacking to fit the one-halo and two-halo amplitudes in our stacking data. Details on modeling galaxy clustering in the stacking signals are further described in Sec.~\ref{S:Model_Signal}.

\section{Stacking}\label{S:stacking}
\subsection{Sub-pixel Stacking}\label{S:Subpix_Stacking}
CIBER imager pixels under-sample the PSF and therefore the surface brightness profile of individual sources is poorly resolved. However, given external source catalogs with high astrometric accuracy, we can stack on a sub-pixel basis and reconstruct the average source profile at scales finer than the native pixel size. This ``sub-pixel stacking'' technique has been used in previous CIBER imager analyses \citep{2013ApJS..207...32B,2014Sci...346..732Z} and further investigated recently in the context of optimal photometry \citep{2021ApJS..252...24S}. We summarize the sub-pixel stacking procedure as follows:
\begin{enumerate}
    \item Select a list of stacking target sources from external catalogs;
    \item Re-grid each pixel into $N_{\rm sub}\times N_{\rm sub}$ sub-pixels (we use $N_{\rm sub}=10$ in this work). The intensities of all sub-pixels are assigned to the same value as the native pixel without interpolation;
    \item For each source, unmask pixels associated with its source mask. Pixels masked due to nearby sources or from the instrument mask remain masked;
    \item Crop an $N_{\rm size}\times N_{\rm size}$ (at sub-pixel resolution) stamp centered on the target source. We choose $N_{\rm size}=2401$ in this work, which corresponds to a $28'\times 28'$ stamp;
    \item Repeat steps 3 and 4 for all target sources, average the stamps, and return the final stacked 2D image $\Sigma_{\rm stack}(\mathbf{r})$.
\end{enumerate}

The stacked profile $\Sigma_{\rm stack}$ is a convolution of the intrinsic source profile, $\Sigma_{\rm src}$, the instrument PSF ($PSF_{\rm instr}$)\footnote{Instrument PSF includes all effects from the optics, detector array, and pointing jitter during the integration.}, and the pixel function $PSF_{\rm pix}$:
\begin{equation}\label{E:Sigma_stack}
\begin{split}
\Sigma_{\rm stack}(\mathbf{r}) =& \left[\Sigma_{\rm src}(\mathbf{r})\circledast PSF_{\rm instr}(\mathbf{r})\right]\circledast PSF_{\rm pix}(\mathbf{r})\\
=&\Sigma_{\rm src}(\mathbf{r})\circledast PSF_{\rm stack}(\mathbf{r}),
\end{split}
\end{equation}
where $\mathbf{r}=(x, y)$ is a two-dimensional sub-pixel coordinate system with its origin at the stack center.  We define the effective PSF as $PSF_{\rm stack} (\mathbf{r}) \equiv PSF_{\rm instr}(\mathbf{r})\circledast PSF_{\rm pix}(\mathbf{r})$. The pixel function accounts for the fact that sub-pixels retain the value of the original pixels, which is a convolution effect. The pixel function is a matrix with each element proportional to the counts where the sub-pixel and the center sub-pixel that contains the source are within the same native pixel. The position of the center sub-pixel within the native pixel is a uniform probability distribution, and therefore when stacking on a large number of sources, the pixel function converges to the analytic form \citep{2021ApJS..252...24S}:
\begin{eqnarray}\label{E:PSF_pix}
PSF_{\rm pix} (\mathbf{r})  = 
 \left\{
\begin{array}{cc}
(N_{\rm sub} - x) (N_{\rm sub} - y) & \\
\quad \quad \quad \quad \quad \quad \text{if }|x|, |y| < N_{\rm sub} \\
0 \quad \quad \text{otherwise}
\end{array}\right.
\end{eqnarray}
As a practical matter, $PSF_{\rm pix}$ can be determined through simulations.
$PSF_{\rm stack}(\mathbf{r})$ can be measured by stacking stars in the field, where $\Sigma_{\rm src}(\mathbf{r})$ is a delta function, so $\Sigma_{\rm stack}(\mathbf{r})=PSF_{\rm stack}(\mathbf{r})$. Note that the expression in the second line of Eq.~\ref{E:Sigma_stack} implies that the intrinsic profile $\Sigma_{\rm src}(\mathbf{r})$ can be obtained from the stacked profile $\Sigma_{\rm stack}(\mathbf{r})$ with the knowledge of $PSF_{\rm stack}(\mathbf{r})$, instead of determining $PSF_{\rm instr}(\mathbf{r})$.

We perform stacking and PSF modeling separately for each field, since $PSF_{\rm instr}$ is slightly different across the fields due to the varying pointing performance of the altitude control system during each integration (see top panel of Fig.~\ref{F:PSF_RLD}). After obtaining the 2D stacked images, we bin them into 25 logarithmically spaced 1D radial bins. Within each bin, the number of stacked images on each sub-pixel is used for weighting when calculating the average profile in each radial bin. Note that the weight is not the same across sub-pixels since the masks are different for each stacked image.

\subsection{Covariance Matrix of Stacking Profile}\label{S:covariance}
The covariance matrix of the binned 1D radial stacked profile is calculated with a jackknife resampling technique. For each stack, we split sources into $N_{\rm J}=64$ sub-groups based on their spatial coordinates in the image. The CIBER imager arrays have 1024$\times$1024 pixels, and thus each sub-group corresponds to sources in a $128\times 128$ pixel sub-region on the array. The radial profile of the $k$th jackknife sample, $\Sigma^k_{\rm stack}$, is obtained from stacking on sources in all the other sub-regions, and then the covariance matrix between radial bin ($r_i$, $r_j$) is given by
\begin{equation}\label{E:C_stack}
\begin{split}
&C_{\rm stack}(r_i, r_j) = \frac{N_{\rm J}-1}{N_{\rm J}}
\sum_{k=1}^{N_{\rm J}}\left [\Delta\Sigma^k_{\rm stack}(r_i)\cdot\Delta\Sigma^k_{\rm stack}(r_j)  \right ]\\
&\Delta\Sigma^k_{\rm stack}(r_i) \equiv \Sigma^k_{\rm stack}(r_i) - \Sigma_{\rm stack}(r_i)\\
&\Delta\Sigma^k_{\rm stack}(r_j) \equiv \Sigma^k_{\rm stack}(r_j) - \Sigma_{\rm stack}(r_j),
\end{split}
\end{equation}
where $\Sigma_{\rm stack}$ is the average stacked profile of all of the sub-regions. 

One of our galaxy stacking samples (mag bin \# 1 in Sec.~\ref{S:Source_Selection}) has a small number of sources ($\ll 64$ for each field), which makes the covariance estimation from the jackknife method unstable. Therefore we perform bootstrap resampling with $N_B=1000$ realizations to calculate the covariance for this case. In this bootstrap, we obtain the radial profile of the $k$th bootstrap sample, $\Sigma_{\rm stack}^{k}$, by stacking the same number of sources as the original sample, but the sources are randomly selected from the original sample with replacement. The covariance matrix is then given by
\begin{equation}
C_{\rm stack}(r_i, r_j) = \frac{1}{N_{\rm B}-1}
\sum_{k=1}^{N_{\rm B}}\left [\Delta\Sigma^k_{\rm stack}(r_i)\cdot\Delta\Sigma^k_{\rm stack}(r_j)  \right ].
\end{equation}
In all the other cases, the covariance is derived from jackknife instead of bootstrap resampling since it is numerically expensive to perform a sufficient number of bootstrap realizations given that we have hundreds or thousands of galaxies per field in each stack. We assign galaxies to sub-groups by their spatial positions instead of randomly grouping them to account for large-scale spatial fluctuations.

The first few radial bins within the CIBER $7''$ native pixel are highly correlated since all the sub-pixels are assigned to the same value as the native pixel. We also find a high correlation on large angular scales, as the stacking signal is dominated by large-scale spatial variations.

\section{PSF Modeling}\label{S:PSF}
An accurate model for the PSF is essential for quantifying the galaxy extension from stacking images.  As stars are point sources on the sky, we measure the PSF of each field by stacking stars in the same CIBER field. The radial profile of star stacks gives $PSF_{\rm stack}$ (Eq.~\ref{E:Sigma_stack}), which accounts for all effects that distribute the light from a point source to the stacked profile, including spreading by the instrument optical system and detectors, pointing instability during integration, astrometry uncertainties, and the pixel function $PSF_{\rm pix}$. Since we use bright stars in the CIBER fields to model the PSF, the uncertainty on the PSF is subdominant to our galaxy stacked profiles.

\def\stackPSF{$PSF_{\rm stack}$}
\subsection{Modeling \stackPSF}
Infrared detectors have a brightness-dependent PSF, the so-called ``brighter-fatter effect'' \citep{2020PASP..132a4501H}. This nonlinearity makes brighter point sources appear broader on the detector array than fainter ones. To model $PSF_{\rm stack}$ robustly on both small and large scales, we construct an overall star profile from three brightness bins. For the core region ($r<22''$), we stack $13<m_{\rm 1.1}<14$ sources in the field; for intermediate scales, $22''<r<40''$, we fit a slope to the stacking profile of $9<m_J^{\rm 2MASS}<10$ sources; for outer radii, we fit another slope to the stacking profile of the brightest $4<m_J^{\rm 2MASS}<9$ sources and connect the two slopes at $r=40''$ ($m_J^{\rm 2MASS}$ is the 2MASS J-band Vega magnitude). The choice of magnitude bins and transition radii minimizes the error on all scales. At small radii, using faint stars avoids detector nonlinearity and at large radii, bright stars provide better sensitivity to the extended PSF. For the intermediate scales, we check that the fitted slope from the three star-stacking profiles ($4<m_J^{\rm 2MASS}<9$, $9<m_J^{\rm 2MASS}<10$, $13<m_{\rm 1.1}<14$) are statistically consistent. The top panel of Fig.~\ref{F:PSF} shows $PSF_{\rm stack}$ from the SWIRE field in the 1.1 $\mu$m band. The top panel of Fig.~\ref{F:PSF_RLD} shows $PSF_{\rm stack}$ in all five fields in both bands. The slight variation across fields is due to the difference in the pointing stability during each integration, but such motion is common to all sources within an integration.

\def\stackPSF{$PSF_{\rm stack}$}
\subsection{Validating \stackPSF}\label{S:PSF_validation}
To validate that our PSF model is applicable to the fainter sources of interest, we perform a consistency test by stacking on stars in the Gaia catalog within the same magnitude range as our stacked galaxy samples ($16<m_{\rm 1.1}<20$) and compare these star-stacking profiles with our $PSF_{\rm stack}$ model.

\begin{figure}[ht!]
\includegraphics[width=\linewidth]{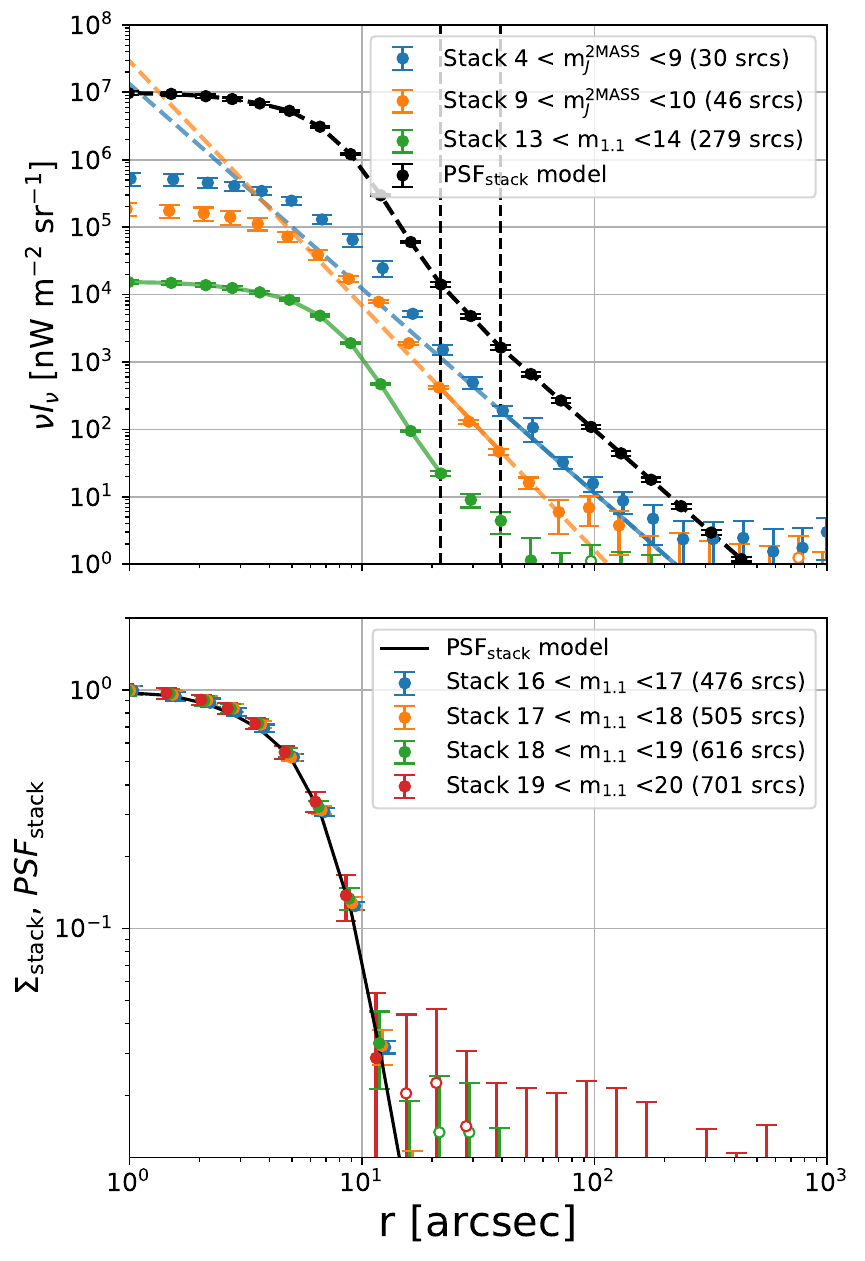}
\caption{\label{F:PSF} We illustrate the process of constructing and validating the $PSF_{\rm stack}(r)$ model, in the 1.1 $\mu$m band SWIRE field. Top: star-stacking profile in three different brightness bins (blue, orange, and green), and the combined $PSF_{\rm stack}(r)$ model (black dashed curve) derived from splicing these three stacking profiles together at the radii marked by the black vertical dashed lines. The black data points show the binned $PSF_{\rm stack}(r)$ and the error bars propagated from their original star stacks. The filled data points and the three colored solid curves are the data used in the $PSF_{\rm stack}(r)$ model. Bottom: comparison of the $PSF_{\rm stack}(r)$ model with the stacking profiles from fainter stars selected from Gaia. The four chosen brightness bins match the ones used in galaxy stacking. The $PSF_{\rm stack}(r)$ model agrees closely with the star-stacking profiles as shown in Fig.~\ref{F:PSF_test}.}
\end{figure}

\begin{figure*}[ht!]
\includegraphics[width=\linewidth]{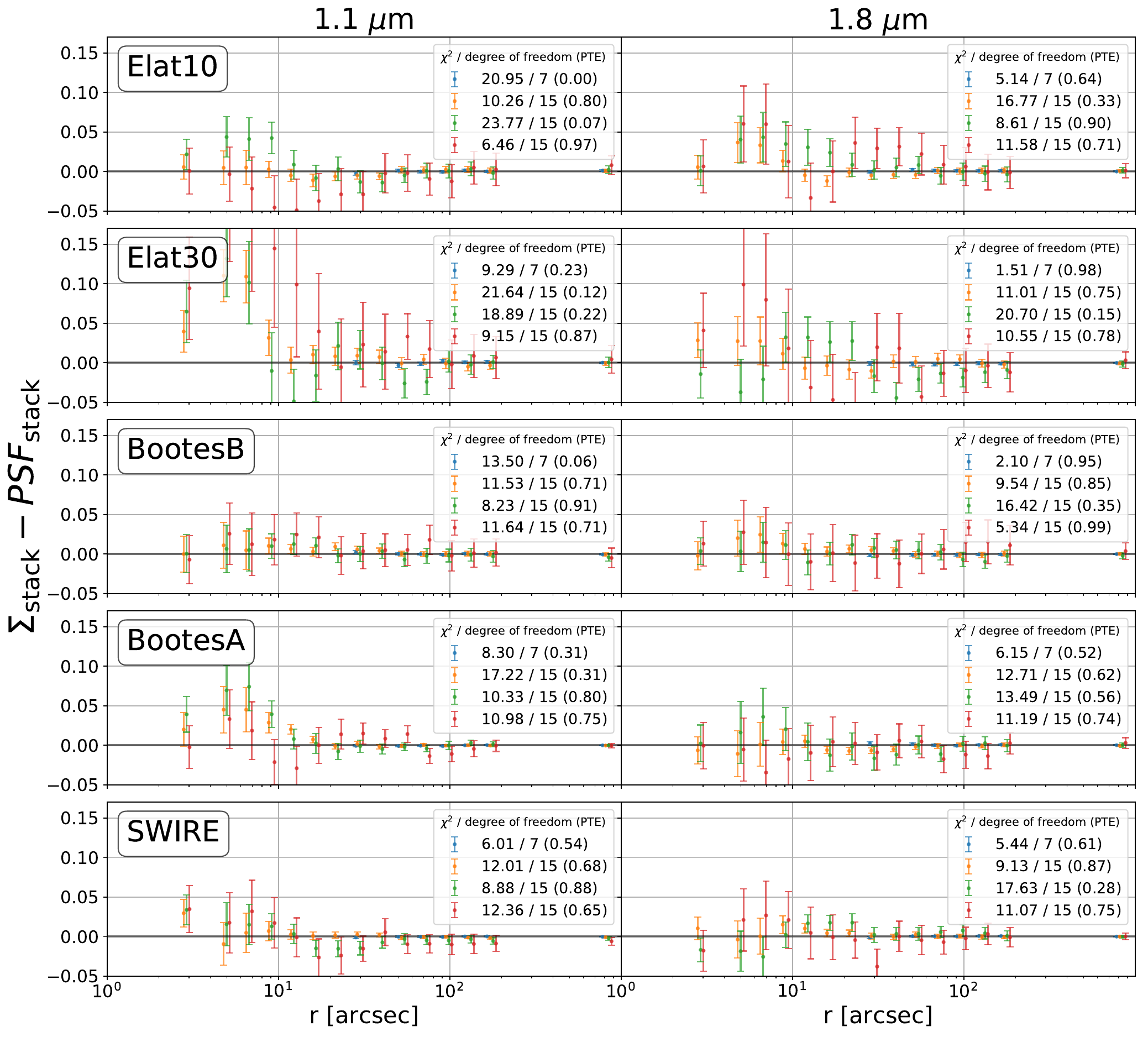}
\caption{\label{F:PSF_test} Difference of the $PSF_{\rm stack}(r)$ model and the star-stacking profiles in all five CIBER fields in the 1.1 $\mu$m (left) and 1.8 $\mu$m (right) bands ($16 < m_{\rm 1.1} <17$ (blue), $17 < m_{\rm 1.1} <18$ (orange), $18 < m_{\rm 1.1} <19$ (green), and $19 < m_{\rm 1.1} <20$ (red)). The $\chi^2$ values and their corresponding PTE given in the legend are consistent with the model. The degrees of freedom for each case are simply the number of radial bins. Open circles in the top and middle panels represent negative data points.}
\end{figure*}

To get a clean star sample free of galaxies, we apply the following criteria for selecting stars from Gaia: 
\begin{enumerate}
    \item The source has a parallax measurement $> 2\times 10^{-4}$ mas (i.e., distance $<$ 5 kpc);
    \item No astrometric excess noise is reported in the Gaia catalog ({\tt\string astrometric\_excess\_noise} $=0$). Large astrometric excess noise implies the source might be extended rather than a point source;
    \item No SDSS galaxies within $0.7^{''}$ (sub-pixel grid size) radius around the source;
    \item We classify SDSS stars and galaxies using 10 pairs of magnitude differences between the 5 Pan-STARRS photometric magnitudes ($g$, $r$, $i$, $z$, and $y$ bands), rejecting sources if they are classified as galaxies by our trained model. 
\end{enumerate}
After selecting stars with the above conditions from the the Gaia catalog, we stack them in four equally-spaced magnitude bins between $16<m_{\rm 1.1}<20$, and compare their stacking profile with the $PSF_{\rm stack}$ model. These stars span the same brightness range used for galaxy stacking. We down-sample the original 25 radial bins to 15 bins (7 bins for $16<m_{\rm 1.1}<17$ case), following the same binning used for the galaxy stacking profile (Sec.~\ref{S:excess}). The results in the 1.1 $\mu$m band SWIRE field are shown on the bottom panel of Fig.~\ref{F:PSF}. In Fig.~\ref{F:PSF_test} we show the difference of Gaia star stacks and the $PSF_{\rm stack}$ model. The errors are propagated from the covariance of the $PSF_{\rm stack}$ model and Gaia star stacks. We also show the $\chi^2$ values and the corresponding probability to exceed (PTE) on all five CIBER fields in both bands. The PSF model shows excellent agreement with the star stacks.

\def\instrPSF{$PSF_{\rm instr}$}
\subsection{Modeling \instrPSF}\label{S:PSF_instr}
Although knowledge of the instrument PSF is not required for reconstructing the source profile $\Sigma_{\rm src}$ from the stacking profile $\Sigma_{\rm stack}$, $PSF_{\rm instr}$ is still needed when we model the clustering signal from a simulated catalog (Sec.~\ref{S:Model_Signal}), where we make mock galaxy images using the CIBER PSF and pixel gridding. $PSF_{\rm instr}$ is also useful for determining the masking radius for bright sources (Sec.~\ref{S:Source_Mask}).

$PSF_{\rm instr}$ is modeled as follows: first, we deconvolve $PSF_{\rm pix}(\mathbf{r})$ (Eq.~\ref{E:PSF_pix}) from the $PSF_{\rm stack}(\mathbf{r})$ model with 10 iterations of the Richardson--Lucy deconvolution algorithm \citep{1972JOSA...62...55R,1974AJ.....79..745L}. 
The deconvolution is unstable at large radii due to noise fluctuations. To get a smooth model for $PSF_{\rm instr}$, we fit a $\beta$ model \citep{1978A&A....70..677C} to the 1D profile of the deconvolved image:
\begin{equation}\label{E:PSF}
PSF_{\rm instr}(r) = \left ( 1+\left ( \frac{r}{r_c} \right )^2 \right )^{-3\beta/2}.
\end{equation}
Though not physically motivated, we find the $\beta$ model is a good empirical description of the extended PSF and requires only two free parameters to achieve acceptable goodness of fit for every $PSF_{\rm stack}$.

\begin{figure}[ht!]
\includegraphics[width=\linewidth]{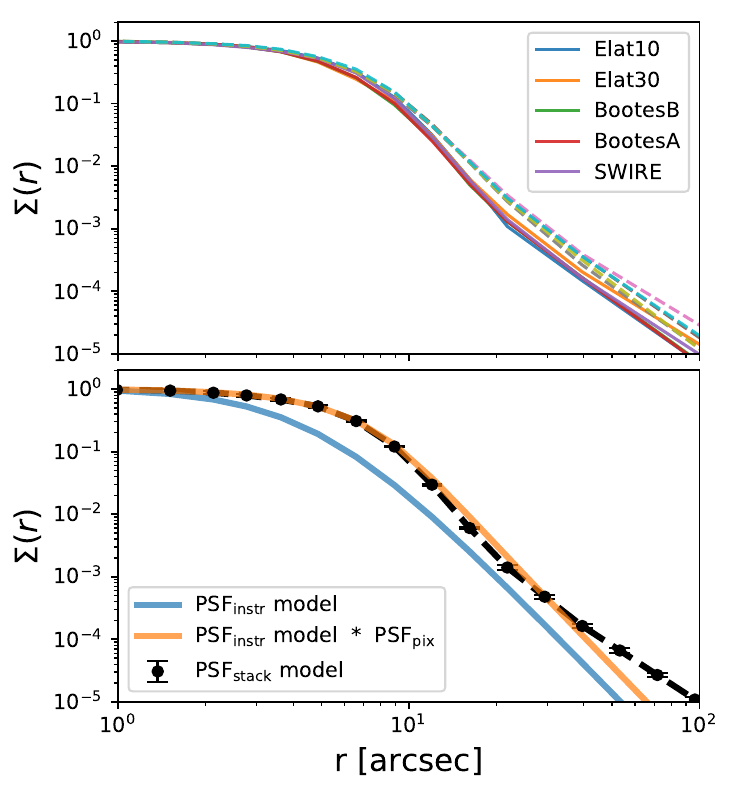}
\caption{\label{F:PSF_RLD} Top: $PSF_{\rm stack}$ model for each of the five fields in the 1.1 $\mu$m (solid)  and 1.8 $\mu$m (dashed) bands. The variation across fields is due to the difference in pointing stability.  Bottom: demonstration of the $PSF_{\rm instr}$ reconstruction process. Black data points show the $PSF_{\rm stack}$ model in the 1.1 $\mu$m band SWIRE field, derived from splicing the star-stacking profile in three different brightness bins (c.f. Fig.~\ref{F:PSF} top panel). The blue line is the $PSF_{\rm instr}$ model derived from fitting a $\beta$ model to $PSF_{\rm stack}$ after deconvolving $PSF_{\rm pix}$ with the Richardson--Lucy  deconvolution algorithm. The orange line shows the convolution of $PSF_{\rm instr}$ with $PSF_{\rm pix}$ matching the $PSF_{\rm stack}$ model, as a consistency check. Our model for $PSF_{\rm instr}$ is in agreement with data for $r\lesssim 30''$. Our analysis is not susceptible to the moderate error at larger radii, as $PSF_{\rm instr}$ is only used for characterizing the clustering signal from nearby galaxies.}
\end{figure}

The bottom panel of Fig.~\ref{F:PSF_RLD} illustrates this procedure in the 1.1 $\mu$m band of the SWIRE field. The $PSF_{\rm stack}$ model, obtained from star stacks in three different brightness bins, matches the $\beta$ model of $PSF_{\rm instr}$ convolved with the pixel function $PSF_{\rm pix}$ (Eq.~\ref{E:PSF_pix}). Our instrument PSF has a size comparable to a pixel (FWHM $\sim 7''$).

\section{Galaxy Stacking}\label{S:Galaxy_Stacking}
We stack galaxies within magnitude ranges $16<m_{\rm 1.1}<20$, divided into several sub-samples spanning $\Delta m_{\rm 1.1}=1$. Our choice of magnitude bins optimizes the SNR on the stacks, giving sufficient sample sizes for each source brightness. 

\subsection{Source Selection Criteria}\label{S:Source_Selection}
The stacking galaxy samples are selected from the SDSS catalog in the CIBER fields. To mitigate systematic effects from confusion, nearby clusters, or misclassified stars in the sample, we reject sources if they meet any of the following criteria:
\begin{itemize}
    \item Sources are not labeled as galaxies in the SDSS catalog, i.e., the ``type'' attribute in the SDSS \texttt{PhotoObj} table is not equal to 3;
    \item Sources are located in the instrument mask;
    \item Other Pan-STARRS sources exist in the same CIBER pixel;
    \item The SDSS photometric redshift is less than 0.15. These criteria prevent nearby galaxies from introducing substantial power on large angular scales that would otherwise mimic the clustering signal;
    \item Sources have nearby Gaia counterparts within $0.7''$, i.e., the size of the sub-pixel used in our stacking. These sources are likely to be stars that are misclassified as galaxies in the SDSS catalog;
    \item Sources are within (1) a 500'' radius of any galaxy cluster in \citet{1958ApJS....3..211A} (Sec.~\ref{S:Nearby_Clusters}); or (2) $R_{200}$ of any galaxy cluster with halo mass $M_h>10^{14}M_\odot$ or redshift $z<0.15$ in the SDSS cluster catalog \citep[][Sec.~\ref{S:Nearby_Clusters}]{2012ApJS..199...34W}. Approximately $10$\% of the sky area in each field is excluded by this condition.
\end{itemize}

The last condition mitigates contamination from nearby clusters along the line of sight since they have structures spanning large angular scales, which will produce spurious large-scale extended signals in the stack. Furthermore, as we do not have information on whether a galaxy in SDSS is a member of a large galaxy cluster, the criteria also exclude cluster members from our stacking sample. Stacking on cluster members introduces extra nonlinear one-halo clustering that can overwhelm the linear two-halo clustering signal on large scales.

To quantify the effect of applying this condition, we generate a mock CIBER map from the MICECAT catalog, implementing the same strategies described above to select sources and stacking on the mock maps to measure the one- and two-halo clustering signals (see Sec.~\ref{S:Model_Signal} for a detailed description of stacking with MICECAT-generated maps). We tested over a range of halo mass and redshift for selecting clusters and found that excluding sources around clusters with $M_h>10^{14}M_\odot$ (or redshift $z<0.15$) can effectively reduce the one-halo clustering signal on large scales without losing a significant number of sources. For example, for the magnitude range of interest in this work (see Sec.~\ref{S:Source_subsamples}), we can reduce the one-halo power by $\sim 3-5\times$ at $100^{''}$ radius just by excluding galaxies near clusters following our criteria.

\subsection{Stacking Sub-samples}\label{S:Source_subsamples}
For the SDSS galaxies within $16<m_{\rm 1.1}<20$ that survive all the selection criteria above, we split the sources into two sets. The first set is based on 1.1 $\mu$m flux in four bins: $16<m_{\rm 1.1}<17$, $17<m_{\rm 1.1}<18$,  $18<m_{\rm 1.1}<19$, and $19<m_{\rm 1.1}<20$. Hereafter, these four bins are named ``mag bin \# 1,'' ``mag bin \# 2,'' ``mag bin \# 3,'' and ``mag bin \# 4,'' respectively. In addition, we also define a ``total stack'' with all $17<m_{\rm 1.1}<20$ sources to achieve better large-scale sensitivity.

The second set is defined by both the 1.1 $\mu$m apparent magnitude $m_{1.1}$ and the absolute magnitude $M_{1.1}$: $M_{1.1} = m_{1.1} - DM(z) + 2.5{\rm log}_{10}(1+z)$, where $DM$ is the distance modulus, using SDSS photometric redshifts. The absolute flux serves as a proxy for galaxy size. Galaxies with comparable absolute flux have similar bolometric luminosity, which is correlated with stellar mass, star formation rate, etc. We use these samples to explore the dependence of our results on different galaxy properties. Since the sets approximately correspond to three higher and two lower stellar mass populations, with different redshift distributions, we call them ``high-M/low-z,'' ``high-M/med-z,'' ``high-M/high-z,'' ``low-M/low-z,'' and ``low-M/med-z.'' 

In the SWIRE field, we have additional information from a photometric redshift catalog \citep{2013MNRAS.428.1958R} based on an SED fit to each galaxy. As the stacked samples from each field are selected with the same criteria, we can assume the galaxy property distributions in the SWIRE field are the same as other fields, and thus infer the stellar mass distribution over all five fields. The log $M_*$ column in Table~\ref{T:mag_bins} lists the median and 68\% interval stellar mass in the SWIRE field samples from the \citet{2013MNRAS.428.1958R} catalog. The stellar masses of our samples span from $\sim 10^{10.5}$ to $10^{12}M_\odot$, i.e., $\sim L_*$ galaxies at this redshift \citep{2013ApJ...777...18M}. In addition, with the stellar mass distribution, we infer the host halo mass of our samples using the mean stellar-to-halo mass relation given by \citet{2015MNRAS.454.1161Z}, which connects the halo mass to stellar mass with galaxy clustering and lensing measurements. We also derive the corresponding virial radius, $R_{\rm 200}$ (in physical and angular units), in Table~\ref{T:mag_bins}. The virial radius is calculated from $R_{\rm 200}=[3M_h/(4\pi\cdot 200\rho_c)]^{1/3}$, where $\rho_c$ is the critical density.

\begin{figure*}[ht!]
\includegraphics[width=\linewidth]{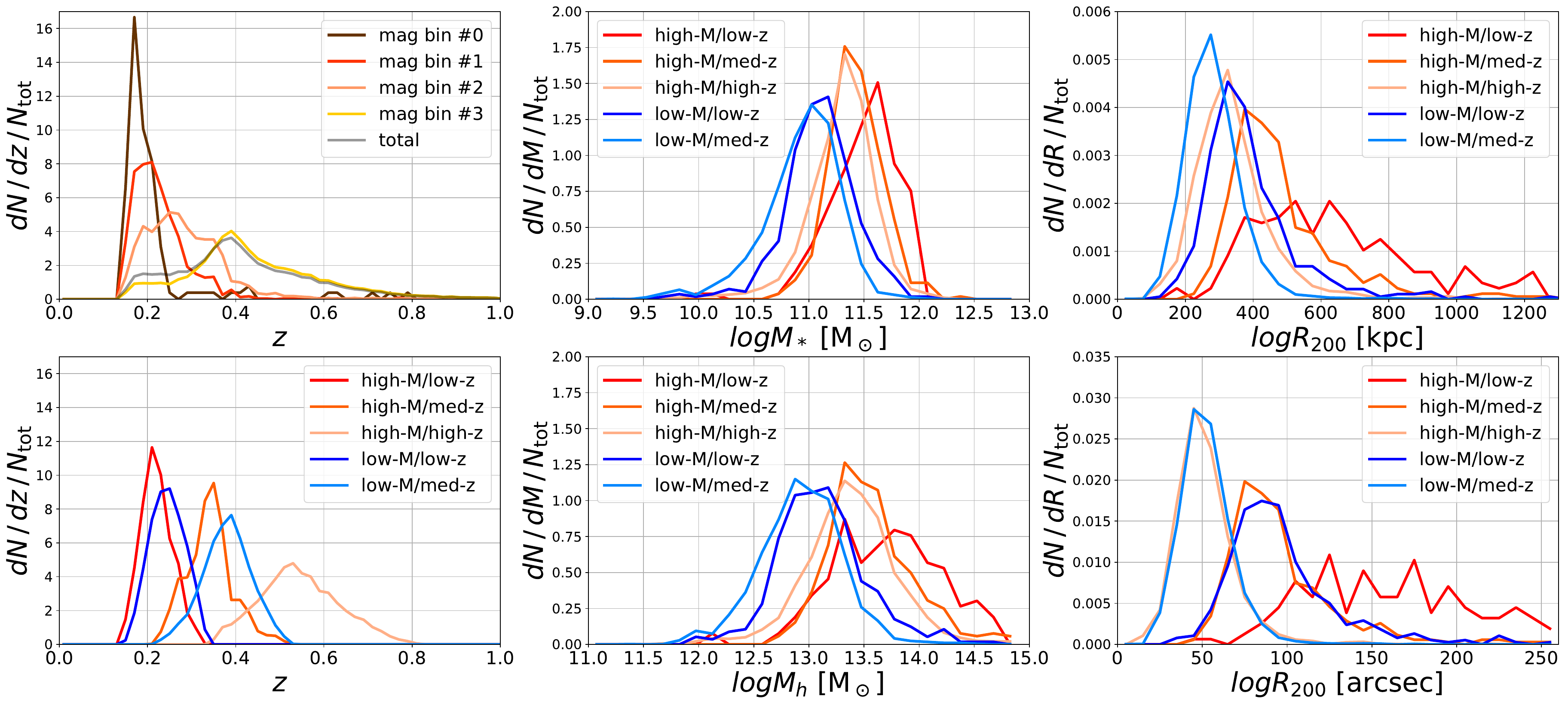}
\caption{\label{F:bins_hist} Left: redshift distributions of the 10 galaxy sub-samples used for stacking. The redshifts are derived from SDSS photometry. Middle top: stellar mass distributions for the 5 apparent and absolute magnitude selected bins. The stellar masses are inferred from \citet{2013MNRAS.428.1958R} for the SWIRE field. Middle bottom: halo mass distributions in 5 apparent and absolute magnitude selected bins, modeled by applying the stellar-to-halo mass relation from \citet{2015MNRAS.454.1161Z}. Right: distributions of virial radius in physical (top) and observed angular (bottom) units. For visualization purposes, all curves are normalized by the total number of sources in each sub-sample ($N_{\rm tot}$).}
\end{figure*}

We note that by selecting galaxies based on absolute or apparent fluxes, our samples will include both central and satellite galaxies. We infer the fraction of central galaxies, $f_{\rm cen}$, in each sub-sample from MICECAT by applying the same selection criteria from a MICECAT simulation (i.e., observed magnitude, absolute magnitude and redshift cuts and excluding sources close to nearby clusters). The distribution of redshift, stellar mass, halo mass, virial radius, and $f_{\rm cen}$ of our sub-samples is summarized in Fig.~\ref{F:bins_hist} and Table~\ref{T:mag_bins}.

\begin{deluxetable*}{c|ccc||cccc|c}[t]
\tablenum{2}
\label{T:mag_bins}
\tablecaption{Summary of the properties on each stacked galaxy sub-sample with the $+/-$ values indicating the 68\% interval ranges.}
\tablewidth{0pt}
\tablehead{
\colhead{Name} & Selection Criteria & \colhead{$N_{\rm gal}$} & \colhead{$z$} & \colhead{log $M_*$ ($M_\odot$)} & \colhead{log $M_h$ (M$_\odot$)} & \colhead{$R_{\rm 200}$ (kpc)} & \colhead{$R_{\rm 200}$ (arcsec)}& \colhead{$f_{\rm cen}$}
}
\startdata
mag bin \#1 & $16<m_{\rm 1.1}<17$ & 129 & 0.18$^{+0.04}_{-0.02}$ & 11.6$^{+0.3}_{-0.3}$ & 13.8$^{+0.5}_{-0.4}$ & 679$^{+325}_{-181}$ & 215$^{+103}_{-57}$ & 0.65\\
mag bin \#2 & $17<m_{\rm 1.1}<18$ & 1173 & 0.21$^{+0.07}_{-0.04}$ & 11.5$^{+0.3}_{-0.4}$ & 13.7$^{+0.6}_{-0.6}$ & 584$^{+357}_{-215}$ & 163$^{+100}_{-60}$ & 0.67\\
mag bin \#3 & $18<m_{\rm 1.1}<19$ & 3465 & 0.27$^{+0.09}_{-0.07}$ & 11.2$^{+0.4}_{-0.3}$ & 13.3$^{+0.5}_{-0.4}$ & 401$^{+178}_{-116}$ & 94$^{+42}_{-27}$ & 0.62\\
mag bin \#4 & $19<m_{\rm 1.1}<20$ & 31157 & 0.42$^{+0.17}_{-0.11}$ & 11.1$^{+0.3}_{-0.5}$ & 13.0$^{+0.5}_{-0.5}$ & 285$^{+127}_{-86}$ & 50$^{+22}_{-15}$ & 0.63\\\hline
total & $17<m_{\rm 1.1}<20$ & 35795 & 0.40$^{+0.17}_{-0.14}$ & 11.1$^{+0.3}_{-0.4}$ & 13.1$^{+0.5}_{-0.5}$ & 302$^{+135}_{-93}$ & 55$^{+24}_{-17}$ & 0.63\\\hline
high-M/low-z & \makecell{$17<m_{\rm 1.1}<18$\\$-23<M_{\rm 1.1}<-22$} & 743 & 0.22$^{+0.04}_{-0.03}$ & 11.6$^{+0.2}_{-0.4}$ & 13.7$^{+0.5}_{-0.5}$ & 608$^{+266}_{-201}$ & 168$^{+73}_{-55}$ & 0.66\\
high-M/med-z & \makecell{$18<m_{\rm 1.1}<19$\\$-23<M_{\rm 1.1}<-22$} & 1274 & 0.34$^{+0.05}_{-0.05}$ & 11.4$^{+0.3}_{-0.2}$ & 13.5$^{+0.4}_{-0.3}$ & 447$^{+157}_{-94}$ & 89$^{+31}_{-19}$ & 0.62\\
high-M/high-z & \makecell{$19<m_{\rm 1.1}<20$\\$-23<M_{\rm 1.1}<-22$} & 10916 & 0.54$^{+0.10}_{-0.09}$ & 11.3$^{+0.3}_{-0.3}$ & 13.4$^{+0.3}_{-0.4}$ & 325$^{+100}_{-82}$ & 50$^{+15}_{-13}$ & 0.66\\
low-M/low-z & \makecell{$18<m_{\rm 1.1}<19$\\$-22<M_{\rm 1.1}<-21$} & 1645 & 0.24$^{+0.05}_{-0.03}$ & 11.1$^{+0.3}_{-0.2}$ & 13.1$^{+0.4}_{-0.3}$ & 359$^{+129}_{-78}$ & 90$^{+33}_{-20}$ & 0.57\\
low-M/med-z & \makecell{$19<m_{\rm 1.1}<20$\\$-22<M_{\rm 1.1}<-21$} & 14730 & 0.38$^{+0.05}_{-0.05}$ & 11.0$^{+0.2}_{-0.4}$ & 12.9$^{+0.3}_{-0.4}$ & 275$^{+78}_{-67}$ & 51$^{+15}_{-13}$ & 0.58
\enddata
\tablecomments{$N_{\rm gal}$ is the total number of galaxies across five CIBER fields in each sub-sample and the redshifts $z$ are derived from SDSS photometry. The quantities on the left side of the double vertical line are derived from a partial set of samples or external catalogs for the sources used in stacks. We infer $M_*$ by matching SWIRE field sources to the catalog from \cite{2013MNRAS.428.1958R}, assuming the same $M_*$ distribution applies to the other four fields. The halo mass and the virial radius are derived with the stellar-to-halo mass relation from \citet{2015MNRAS.454.1161Z}. The fraction of central galaxies ($f_{\rm cen}$) is derived by applying the same cuts to a simulated catalog from MICECAT.}
\end{deluxetable*}

\subsection{Galaxy Stacking Profile}\label{S:gal_stack}

\begin{figure}[ht!]
\includegraphics[width=\linewidth]{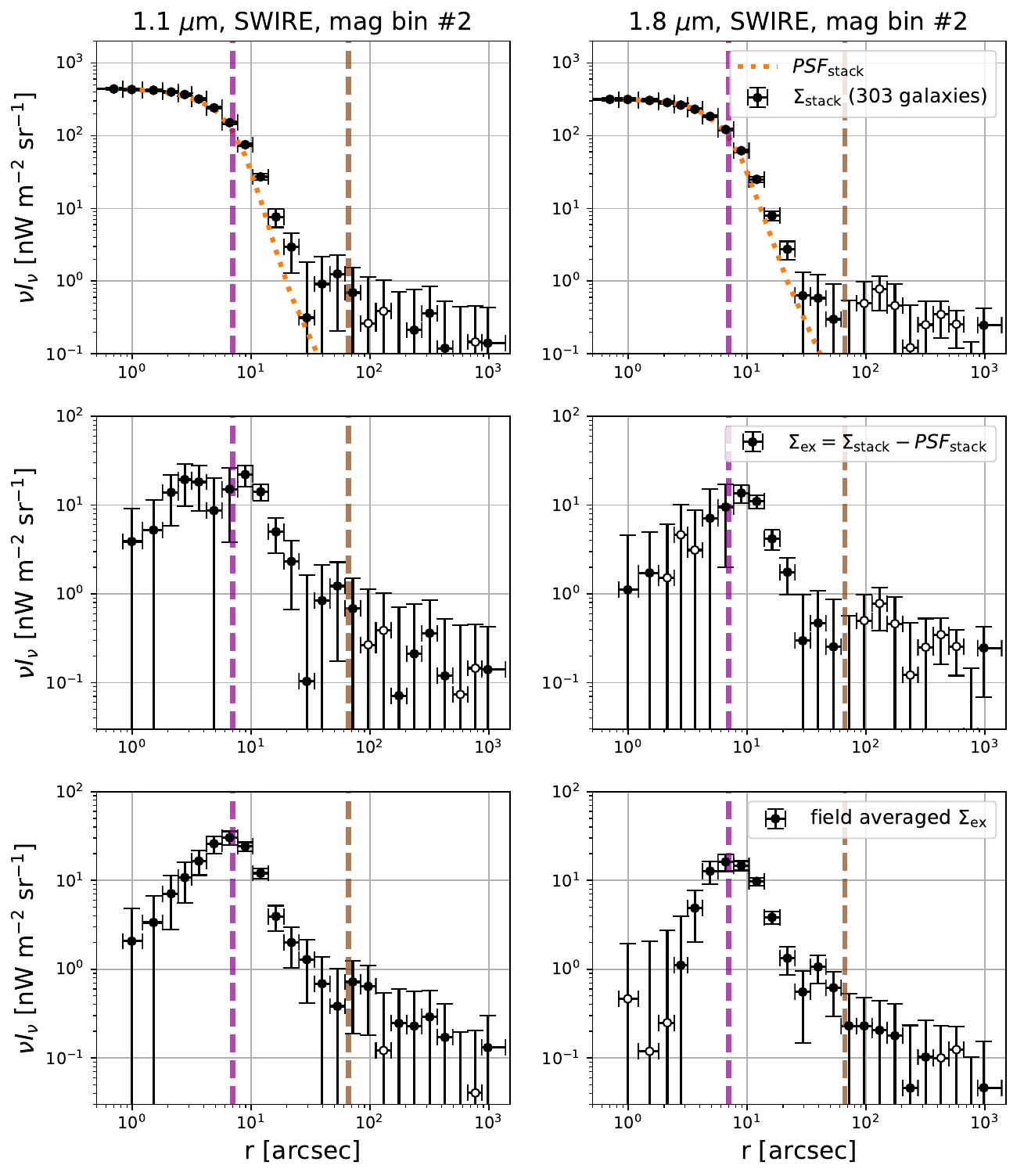}
\caption{\label{F:stack_profile} Stacked galaxy radial profile from the SWIRE field mag bin \#2 in the 1.1 (left) and 1.8 $\mu$m (right) bands. Top: galaxy stacked profile $\Sigma_{\rm stack}$ (black) and $PSF_{\rm stack}$ model (orange dashed), scaled to match the innermost radial bin of $\Sigma_{\rm stack}$. The error bars give the diagonal element of the covariance matrix derived by the jackknife method (described in Sec.~\ref{E:C_stack}). Middle: the excess profile ($\Sigma_{\rm ex}$, Eq.~\ref{E:Sigma_ex}) for the case shown in the top row. The excess is defined as the difference between the galaxy stacked profile and the $PSF_{\rm stack}$ model, i.e., the difference of the black data from the orange curve in the top row. Bottom: the field-averaged excess profile $\Sigma_{\rm ex}$ for mag bin \#2, derived from the weighted average of the excess profile in the five individual fields. The improved sensitivity from combining fields can be seen compared to the middle row. The purple and brown dashed lines mark the pixel size and the median $R_{\rm 200}$ values inferred from MICECAT, respectively. Open circles in all the plots represent negative values.}
\end{figure}

We calculate 1D radial profiles from galaxy stacks by averaging pixels in concentric annuli, as shown in Fig.~\ref{F:stack_profile} and Fig.~\ref{F:stack_profile_fit}. For comparison, we also plot the expected profile of stacked point sources, $PSF_{\rm stack}$, scaled to match the first radial bin of the stacked galaxy profile. In all cases, the galaxy profiles are clearly broader than the $PSF_{\rm stack}$ profile.

\subsection{Excess Profile}\label{S:excess}
We define an ``excess profile'' $\Sigma_{\rm ex}(r)$ as follows:
\begin{equation}\label{E:Sigma_ex}
\Sigma_{\rm ex}(r) = \Sigma_{\rm stack}(r) - A\cdot PSF_{\rm stack}(r),
\end{equation}
where the normalization factor $A$ is chosen such that $PSF_{\rm stack}$ matches $\Sigma_{\rm stack}$ at the innermost radial bin $r_1$, and thus by construction, $\Sigma_{\rm ex}(r_1)=0$, and $A \equiv \Sigma_{\rm stack}(r_1) / PSF_{\rm stack}(r_1)$.

Since the excess profile is fixed at $r_1$, the uncertainties on the galaxy profile and the PSF profile at $r_1$ have to be accounted for by propagating this error to the other radial bins, and thus the excess profile covariance is given by
\begin{equation}\label{E:C_ex}
C_{\rm ex} = \Sigma_{\rm stack}(r_1)^2\left [C_{\rm norm}\left ( C_{\rm stack}\right ) + C_{\rm norm}\left ( C_{\rm PSF}\right )  \right ],
\end{equation}
where $C_{\rm PSF}$ and $C_{\rm stack}$ are the covariance of $PSF_{\rm stack}$ and $\Sigma_{\rm stack}$, respectively, and
\begin{equation}\label{E:Cnorm}
\begin{split}
C_{\rm norm}&\left ( C, \frac{\Sigma_{\rm stack}(r_i)}{\Sigma_{\rm stack}(r_1)},\frac{\Sigma_{\rm stack}(r_j)}{\Sigma_{\rm stack}(r_1)}\right ) \\
&= \frac{\Sigma_{\rm stack}(r_i)\Sigma_{\rm stack}(r_j)}{\Sigma_{\rm stack}(r_1)^2}\cdot\\
&\left[\frac{C(r_i,r_j)}{\Sigma_{\rm stack}(r_i)\Sigma_{\rm stack}(r_j)}-\frac{C(r_i,r_1)}{\Sigma_{\rm stack}(r_i)\Sigma_{\rm stack}(r_1)}\right.\\
&\left.-\frac{C(r_j,r_1)}{\Sigma_{\rm stack}(r_j)\Sigma_{\rm stack}(r_1)}+\frac{C(r_1,r_1)}{\Sigma_{\rm stack}(r_1)^2}  \right ]
\end{split}
\end{equation}
is the covariance for the normalized profile that follows from the product rule for derivatives.


To fit a model to the measured $\Sigma_{\rm ex}$, we also need the inverse of $C_{\rm ex}$. However, $C_{\rm ex}$ is close to singular since our radial bins are highly correlated. Therefore, we reduce the original 25 radial bins to 15 bins by combining highly correlated bins in the inner and outer regions\footnote{Mag bin \# 1 is down-sampled to 7 radial bins as its degree of freedom is limited by the small number of stacked sources.}. After this down-sampling, we derive the inverse covariance estimator by
\begin{equation}\label{E:C_inv}
C_{\rm ex}^{-1} = \frac{N_{\rm J}-N_{\rm bin}-2}{N_{\rm J}-1}C_{\rm ex}^{*-1},
\end{equation}
where $N_{\rm J}=64$, the number of sub-groups used for estimating covariance, and the number of bins $N_{\rm bin}=15$. $C_{\rm ex}^{*-1}$ is the direct inverse of the $C_{\rm ex}$ matrix, and the pre-factor in Eq.~\ref{E:C_inv} de-biases the inverse covariance estimator, as our covariance matrix is derived from our data \citep{2007A&A...464..399H}\footnote{For mag bin \# 1, $N_{\rm bin}=7$, and $N_{\rm J}=64$ is replaced by $N_{\rm B}=1000$ since we use bootstrap resampling method in this case.}.

While we have high sensitivity on the small radial bins of both the galaxy stacked profiles and the PSF model, the $A$ value has minimal dependency on the radius chosen for normalization, and the uncertainty of normalization has been accounted for by the covariance (Eq.~\ref{E:C_ex}), thus our model parameter inference (Sec.~\ref{S:Model_Signal}) does not depend on the definition of the excess profile.

\begin{figure*}[ht!]
\includegraphics[width=\linewidth]{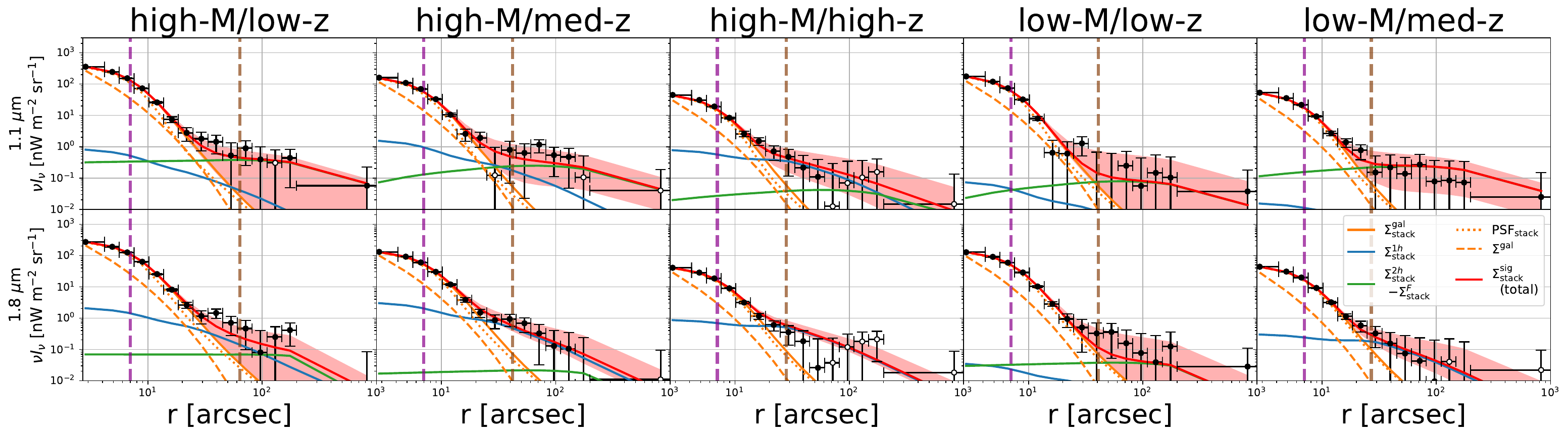}
\caption{\label{F:stack_profile_fit}Stacked profile (black data) of each sub-sample stack averaged over 5 CIBER fields in the 1.1 $\mu$m (top) and 1.8 $\mu$m (bottom) bands. Red lines and shaded regions indicate the median and 68\% confidence interval of the joint fit constrained through MCMC, respectively. The blue, green, and orange solid lines show the best-fit model of the stacked one-halo, two-halo, and galaxy profile term from MCMC. The orange dashed and dotted lines show the best-fit intrinsic galaxy profile $\Sigma^{\rm gal}$ and the $PSF_{\rm stack}$ model. The purple and brown dashed lines mark the pixel size (7$^{''}$) and $R_{200}$ value inferred from MICECAT. Open circles represent negative data points.}
\includegraphics[width=\linewidth]{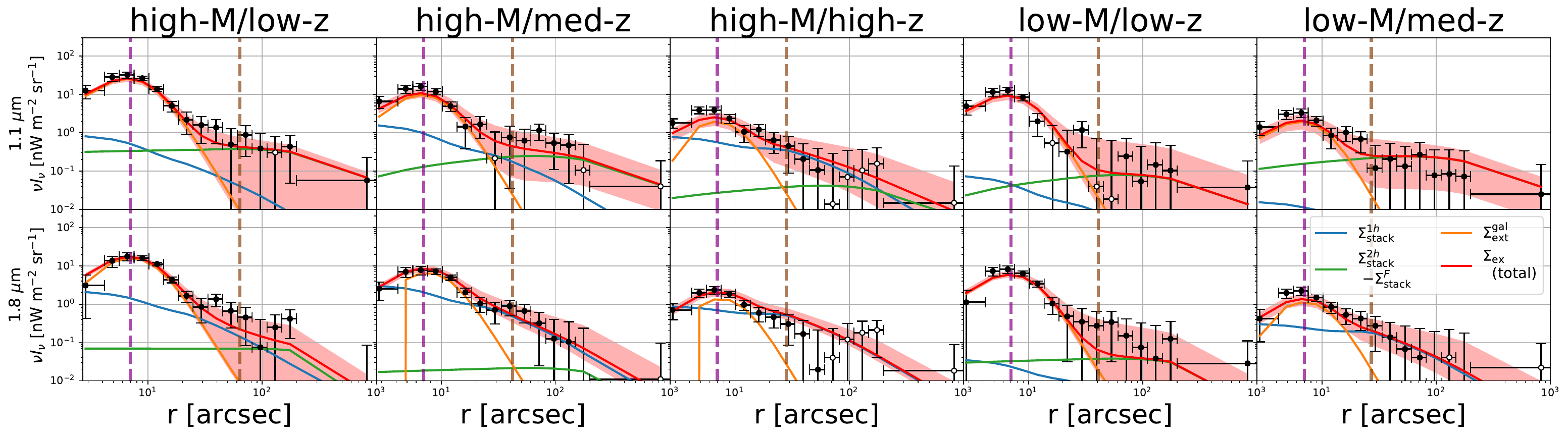}
\caption{\label{F:excess_profile_fit}Measured (black data) and modeled (red) excess profile $\Sigma_{\rm ex}$ (black data) of each case shown in Fig.~\ref{F:stack_profile_fit}. Note the excess profile is defined by the difference of the stacked profile and $PSF_{\rm stacked}$ model (orange dotted line). Other lines are same as the ones shown in Fig.~\ref{F:stack_profile_fit}.}
\end{figure*}

We present field-averaged excess profiles in Fig.~\ref{F:excess_profile_fit}. Note that the field-averaged excess profile is only plotted for visualization purposes since the field-to-field PSF variation must be explicitly accounted for in parameter fitting.

\section{Modeling the Galaxy Profiles}\label{S:Model_Signal}
We model the galaxy profile with three components as follows. We start by decomposing the stacked profile in image space (Sec.~\ref{S:Model_im}), defining fitted profiles (Sec.~\ref{S:Model_stack}), and introducing our model for each component of the stack (Sec.~\ref{S:Model_model}). Finally, we describe the model fitting procedure in Sec.~\ref{S:Model_Fitting}.

\subsection{Components in Image Space}\label{S:Model_im}
The raw CIBER image, $I_{\rm raw}$, can be expressed as\footnote{For clarification, $\mathbf{x}$ denotes 2-D coordinate on CIBER images, and $\mathbf{r}$ represents the coordinate that has origin at the source center, which is used in $PSF_{\rm instr}$ and stacked maps. Since we only consider 1-D radially averaged profile, $\mathbf{r}$ is replaced by 1-D variable ``$r$.''}
\begin{equation}
\begin{split}
I_{\rm raw}(\mathbf{x}) =& \left [ I_{\rm sig}(\mathbf{x}) + I_{\rm LoS}(\mathbf{x}) \right ]\circledast PSF_{\rm instr}(\mathbf{r})\cdot FF(\mathbf{x})\\ 
&+I_{\rm DC}(\mathbf{x}) +I_{\rm n}(\mathbf{x}),
\end{split}
\end{equation}
where $\mathbf{x}$ is the 2D pixel coordinate,  $FF$ is the flat-field gain, $I_{\rm DC}$ is the dark-current map, and $I_{\rm n}$ is the read noise plus photon noise. The sky emission is decomposed into $I_{\rm sig}$ and $I_{\rm LoS}$ terms, where the first term accounts for the signal associated with stacked galaxies and $I_{\rm LoS}$ represents uncorrelated emission from all other sources along the line of sight, including Galactic foregrounds.

After dark-current subtraction and flat-field correction, we retrive $I'_{\rm raw}$:
\begin{equation}\label{E:I_raw}
I'_{\rm raw}(\mathbf{x}) = \left [ I_{\rm sig}(\mathbf{x}) + I_{\rm LoS}(\mathbf{x}) \right ]\circledast PSF_{\rm instr}(\mathbf{r})+I'_{\rm n}(\mathbf{x}),
\end{equation}
where $I'_{\rm n}(\mathbf{x})=I_{\rm n}(\mathbf{x})/FF(\mathbf{x})$, the instrument noise divided by the flat-field response. For simplicity, we ignore the error in the flat-field estimator in Eq.~\ref{E:I_raw}. In practice, the flat-field estimation uncertainties will not bias the stacking results as they are not correlated with individual stacked sources and the effect on the covariance is accounted for by the jackknife method (see Sec.~\ref{S:covariance}). We define the mask $M(\mathbf{x})$ as a binary function set to zero for masked pixels and one otherwise. The filtered map is expressed with $\mathcal{F}\left [ I'_{\rm raw}(\mathbf{x}), M(\mathbf{x}) \right ]$, which is a function of the input map $I'_{\rm raw}(\mathbf{x})$ and mask $M(\mathbf{x})$. As described in Sec.~\ref{S:Image_Filtering}, we choose $\mathcal{F}$ to be a third (1.1 $\mu$m)/fifth (1.8 $\mu$m)-order 2D polynomial function fitted to the masked $I'_{\rm raw}$ map\footnote{Note that the filter map $\mathcal{F}$ can be decomposed into the sum of three filter maps because the polynomial fitting is a linear operation, i.e., given two maps $A(\mathbf{x})$ and $B(\mathbf{x})$, and a mask $M(\mathbf{x})$, $\mathcal{F}\left [ A(\mathbf{x})+B(\mathbf{x}), M(\mathbf{x}) \right ] = \mathcal{F}\left [ A(\mathbf{x}), M(\mathbf{x})\right ]+\mathcal{F}\left [ B(\mathbf{x}), M(\mathbf{x})\right ]$.}. The image used for stacking $I_{\rm map}$ can thus be written as
\begin{equation}\label{E:Imap}
\begin{split}
I_{\rm map}(\mathbf{x}) &= I'_{\rm raw}(\mathbf{x})M(\mathbf{x}) - \mathcal{F}\left [ I'_{\rm raw}(\mathbf{x}), M(\mathbf{x}) \right ]M(\mathbf{x})\\
&= I_{\rm map}^{\rm sig}(\mathbf{x}) + I_{\rm map}^{\rm LoS}(\mathbf{x}) + I_{\rm map}^{\rm n'}(\mathbf{x}),
\end{split}
\end{equation}
where
\begin{equation}\label{E:Isig}
\begin{split}
&I_{\rm map}^{\rm sig}(\mathbf{x}) = \left [ I_{\rm sig}(\mathbf{x})\circledast PSF_{\rm instr}(\mathbf{r}) \right ]M(\mathbf{x}) \\
&- \mathcal{F}\left [I_{\rm sig}(\mathbf{x}) \circledast PSF_{\rm instr}(\mathbf{r}),  M(\mathbf{x}) \right ]M(\mathbf{x}),
\end{split}
\end{equation}

\begin{equation}
\begin{split}
&I_{\rm map}^{\rm LoS}(\mathbf{x}) = \left [ I_{\rm LoS}(\mathbf{x}) \circledast PSF_{\rm instr}(\mathbf{r}) \right ]M(\mathbf{x}) \\
&- \mathcal{F}\left [I_{\rm LoS}(\mathbf{x}) \circledast PSF_{\rm instr}(\mathbf{r}),  M(\mathbf{x}) \right ]M(\mathbf{x}),
\end{split}
\end{equation}
and
\begin{equation}
\begin{split}
&I_{\rm map}^{\rm n'}(\mathbf{x}) = I_{\rm n}'(\mathbf{x})M(\mathbf{x}) \\
&- \mathcal{F}\left [ I_{\rm n}'(\mathbf{x}), M(\mathbf{x}) \right ]M(\mathbf{x}).
\end{split}
\end{equation}

\subsection{Components in the Stack}\label{S:Model_stack}
The stacked profile $\Sigma_{\rm stack}$ can be expressed as the sum of stacks on the three maps in Eq.~\ref{E:Imap}:
\begin{equation}
\Sigma_{\rm stack}(r) = \Sigma_{\rm stack}^{\rm sig}(r) + \Sigma_{\rm stack}^{\rm LoS}(r) + \Sigma_{\rm stack}^{\rm n}(r).
\end{equation}
The last two terms can be ignored in modeling since they are uncorrelated with the stacked sources, so $\left \langle \Sigma_{\rm stack}^{\rm LoS}(r) \right \rangle = \left \langle \Sigma_{\rm stack}^{\rm n}(r) \right \rangle = 0$.

We model the stacked galaxy profile as 
\begin{equation}\label{E:Sigma_stack_sig}
\begin{split}
\Sigma_{\rm stack}^{\rm sig}(r)=\left[\right.\Sigma_{\rm stack}^{\rm gal}(r) + \Sigma_{\rm stack}^{\rm 1h}(r)&+\Sigma_{\rm stack}^{\rm 2h}(r) \left.\right] \\
&- \Sigma_{\rm stack}^{\mathcal{F}}(r),
\end{split}
\end{equation}
where the first three terms are the signal terms and the last term is the filtered signal map in Eq.~\ref{E:Isig}. The galaxy profile term, $\Sigma_{\rm stack}^{\rm gal}$, represents the intrinsic galaxy profile, which includes the galaxy shape and the extended stellar halo. We decompose the galaxy profile term, $\Sigma_{\rm stack}^{\rm gal}$, into ``core'' and ``extended'' parts:
\begin{equation}
\Sigma^{\rm gal}_{\rm stack}(r)= \Sigma_{\rm core}^{\rm gal}(r) + \Sigma_{\rm ext}^{\rm gal}(r),
\end{equation}
where the core component is the integrated emission of the $PSF_{\rm stack}$ fitted to the stacking profile, i.e., the $A\cdot PSF_{\rm stack}$ term in Eq.~\ref{E:Sigma_ex}, and the extended component is the rest of the galaxy emission:
\begin{equation}\label{E:Sigma_core_ext}
\begin{split}
\Sigma_{\rm core}^{\rm gal}(r) =& \Sigma_{\rm stack}(r)-\Sigma_{\rm ex}(r),\\
\Sigma_{\rm ext}^{\rm gal}(r) =& \Sigma_{\rm ex}(r)\\
&-\left[\Sigma_{\rm stack}^{\rm 1h}(r)+\Sigma_{\rm stack}^{\rm 2h}(r)
- \Sigma_{\rm stack}^{\mathcal{F}}(r) \right].
\end{split}
\end{equation}

In addition, galaxy clustering will also contribute to the stacked profile, primarily on large scales. We model clustering with the halo model framework \citep{2002PhR...372....1C}, where large-scale clustering is described by the correlation within (one-halo) and between (two-halo) dark matter halos. $\Sigma_{\rm stack}^{\rm 1h}$ and $\Sigma_{\rm stack}^{\rm 2h}$ represent the profile for one- and two-halo clustering, respectively.

In practice, there is no well-defined boundary between the stellar halo of a galaxy and unbound stars in the dark matter halo, and the definition of IHL (or ICL) varies in the literature. To some degree, the galaxy extension term and the one-halo term each partially comprise stars not bound to individual galaxies in the halo. Since there are different definitions of IHL (or ICL) and the one-halo term in the literature, here we describe how our modeled components are defined.

\begin{figure}[ht!]
\includegraphics[width=\linewidth]{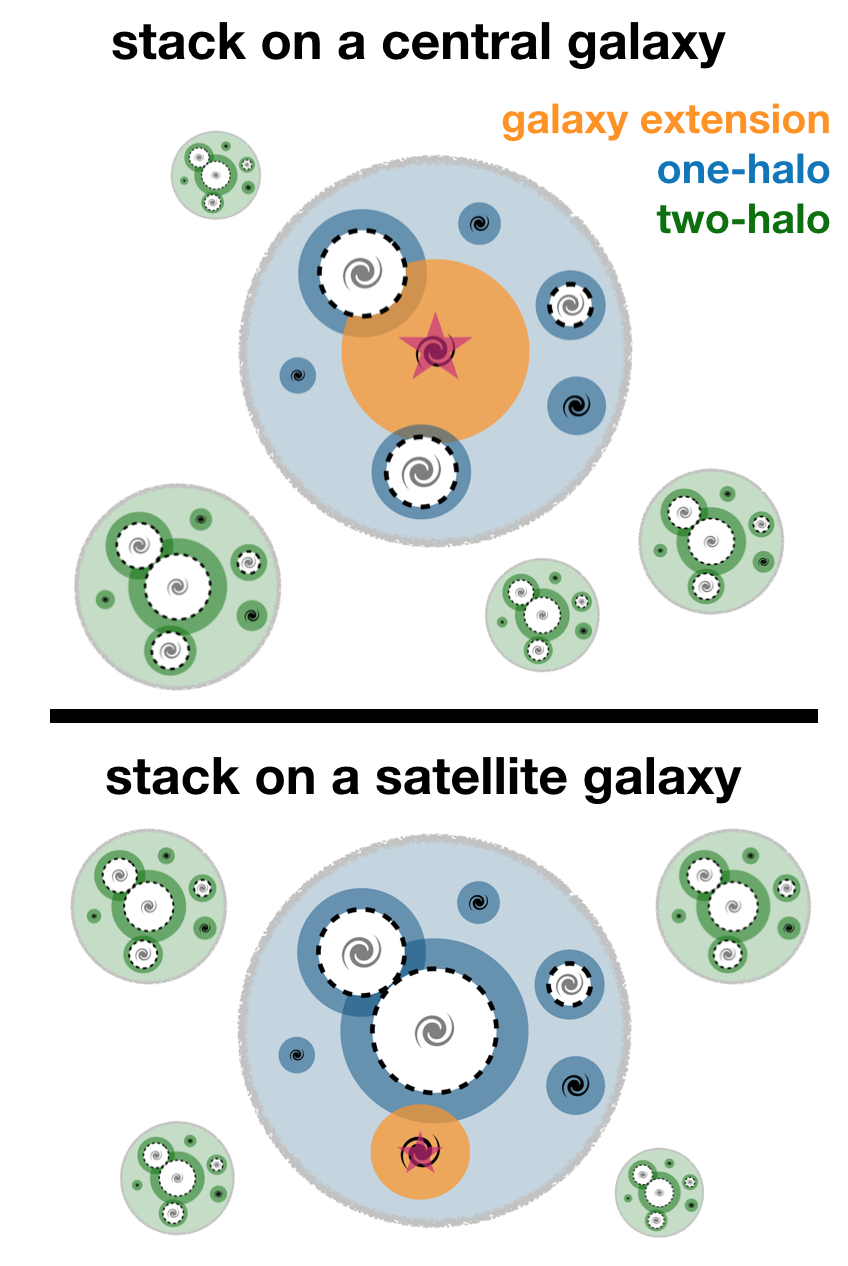}
\caption{\label{F:profile_def} Illustration of the components in our model when stacking on a central (top) or a satellite (bottom) galaxy. The dark regions show the galaxy extensions associated with each galaxy, and the light blue and green regions show diffuse stars in the halos that are not tightly bound to any galaxy. The white parts with black dashed boundaries show the masked regions. The smaller galaxies without masks are fainter than the masking cutoff. The magenta stars and the orange regions show the stacked galaxy and its extension. The blue regions represent the one-halo term, and the green regions show the two-halo term contributed by emission from other halos. When stacking on a central galaxy, the one-halo term includes the satellite galaxy extensions beyond the masking radius as well as faint satellite galaxies and their stellar halos. When stacking on a satellite galaxy, the one-halo term includes the extensions of both the central and the satellite galaxies beyond their masks as well as the fainter satellite galaxies.}
\end{figure}

In our definition, the galaxy extension describes emission associated with each galaxy, whereas the one-halo term accounts for other galaxies, their extensions,  and diffuse stars in the same halo, as illustrated in Fig.~\ref{F:profile_def}. When we stack on a central galaxy, the galaxy extension term accounts for the extended emission around the stacked galaxy, and the one-halo term describes diffuse stars, undetected galaxies, and extension around all the satellite galaxies beyond the masking limit in the same halo. Whereas, when we stack on a satellite galaxy, the galaxy extension term only includes the extended halo around that satellite galaxy, and all the other components are described by the one-halo term. In our sample, we estimate that $\sim 60$\% of stacked galaxies are central galaxies and $\sim 40$\% are satellite galaxies (see Table~\ref{T:mag_bins}). 

\subsection{Modeling the Stacked Galaxy Profile}\label{S:Model_model}
The stacked galaxy profile $\Sigma_{\rm stack}^{\rm gal}(r) = \Sigma^{\rm gal}(r)\circledast PSF_{\rm stack}(r)$, is the intrinsic galaxy profile $\Sigma^{\rm gal}$, including the galaxy shape and the extended stellar halo, convolved with $PSF_{\rm stack}$. Following \cite{2019MNRAS.487.1580W}, we model $\Sigma^{\rm gal}$ with a double Sersic function:
\begin{equation}\label{E:Sersic}
\begin{split}
&\Sigma^{\rm gal}(r) = A^{\rm gal}\left ( 10^{I_{e,1}}{\rm exp}\left \{ -b_{n1}\left [ \left ( r/R_{e,1} \right )^{1/n_1}-1 \right ] \right \}\right.\\
&+\left.10^{I_{e,2}}{\rm exp}\left \{ -b_{n2}\left [ \left ( r/\left (R_{e,1}+R_{e,2}  \right ) \right )^{1/n_2}-1 \right ] \right \} \right ).
\end{split}
\end{equation}
\cite{2019MNRAS.487.1580W} performed a stacking analysis on isolated galaxies from Hyper Suprime-Cam (HSC) images and fitted the stacked profile of their high-concentration samples with this model.  The first term captures the galaxy shape, and the second term models the extended emission. Due to the lack of angular resolution in CIBER data, we are sensitive to the extended profile, and therefore we only vary $R_{e,2}$ to fit our stacked profile. We fix all of the other parameters to the best-fit values given by Table 3 of \cite{2019MNRAS.487.1580W}, although when convolved, the total closely follows the PSF\footnote{In \cite{2019MNRAS.487.1580W}, the values of $R_{e,1}$ and $R_{e,2}$ are reported in terms of $x_{e,1}=R_{e,1}/R_{200}$ and $x_{e,2}=R_{e,2}/R_{200}$. $R_{200}$ is the projected virial radius of the host dark matter halo in angular units and its value for each sub-sample is given in Table~\ref{T:mag_bins}.}.

Our one- and two-halo clustering models, $\Sigma^{\rm 1h}_{\rm stack}$ and $\Sigma^{\rm 2h}_{\rm stack}$, and the filtered signal $\Sigma^{\rm \mathcal{F}}_{\rm stack}$, are constructed from the MICECAT simulation. MICECAT includes central and satellite galaxies of each halo, and each galaxy has a halo ID, enabling us to decouple the one-halo and two-halo contribution in the stacked signal, and thus to take into account the complication that we have both central and satellite galaxies in our samples. We model the one-halo term $\Sigma_{\rm stack}^{\rm 1h}$ from MICECAT using the following steps: 
\begin{enumerate}
    \item Select the stacked target in the catalog using the same selection criteria;
    \item For each target galaxy, generate a source map (using $PSF_{\rm instr}$) for all galaxies residing in the same halo except for the target galaxy;
    \item Generate a source mask using the same prescription as our data;
    \item Stack on the target source position;
    \item Iterate steps (2)--(4) for all target sources.
\end{enumerate}
The derived stacked profile provides our template for the one-halo term, $T_{\rm stack}^{\rm 1h}$. The filtered signal term $\Sigma^{\rm \mathcal{F}}_{\rm stack}$ accounts for the loss of clustering signal from filtering. $\Sigma^{\rm \mathcal{F}}_{\rm stack}$ is the stacked profile on the 2D polynomial filtered map (the second term of Eq.~\ref{E:Isig}), which can be modeled by filtering the simulated map from MICECAT. We model the two-halo term $\Sigma_{\rm stack}^{\rm 2h} - \Sigma_{\rm stack}^{\mathcal{F}}$ after filtering with the following process: 
\begin{enumerate}
    \item Make a CIBER-sized mock image from all the catalog sources with the model $PSF_{\rm instr}$, and mask it with a source mask generated using the same masking process applied to the data;
    \item Fit and subtract a 2D polynomial map to the image;
    \item Select the stacked target in the catalog using the same selection criteria as the real sources;
    \item Perform stacking with the target source, subtracting all galaxies within the same halo to remove the target galaxy and the one-halo contribution;
    \item Iterate on step (4) to derive a stacked profile of the filtered two-halo signal.
\end{enumerate}
The resulting stacked profile,  $T_{\rm stack}^{\rm 2h-\mathcal{F}}$, is a model for $\Sigma^{\rm 2h}_{\rm stack}-\Sigma^{\rm \mathcal{F}}_{\rm stack}$, which provides our template for the two-halo term. This process was performed on 400 realizations with CIBER-sized mock images from MICECAT, and we take the average stacked profile as the one-halo and filtered two-halo templates. As diffuse stars and faint galaxies below the resolution limit of MICECAT will not be accounted for, we assign free amplitudes to the one-halo and two-halo templates, which are then fit to the observed stacked data. Therefore, our three-parameter ($R_{\rm e,2},A_{\rm 1h},A_{\rm 2h}$) model can be written as
\begin{equation}\label{E:Sigma_model}
\begin{split}
\Sigma_{\rm stack}&(r,\left \{ R_{\rm e,2},A_{\rm 1h},A_{\rm 2h} \right \}) \\
&= \Sigma^{\rm gal}(r,\left \{ R_{\rm e,2} \right \})\circledast PSF_{\rm stack}(r) \\
&+ A_{\rm 1h}T_{\rm stack}^{\rm 1h}(r) + A_{\rm 2h}T_{\rm stack}^{\rm 2h-\mathcal{F}}(r).
\end{split}
\end{equation}
We note that the one- and two-halo profiles already include the PSF convolution in our model.

\subsection{Model Fitting}\label{S:Model_Fitting}
For each CIBER field and band, we fit the excess profile Eq.~\ref{E:Sigma_ex}, to a three-parameter model $\Sigma^{\rm m}_{\rm ex}(r,\{ R_{\rm e,2},A_{\rm 1h},A_{\rm 2h}\})$ (Eq.~\ref{E:Sigma_model}) using a Markov Chain Monte Carlo (MCMC). We assume a Gaussian likelihood, which is given by
\begin{equation}
\begin{split}
\chi^2 &=\left ( \Sigma^{\rm d}_{\rm ex}-\Sigma^{\rm m}_{\rm ex} \right )^TC^{-1}_{\rm ex}\left ( \Sigma^{\rm d}_{\rm ex}-\Sigma^{\rm m}_{\rm ex} \right )\\
{\rm ln}\mathcal{L} &= -\frac{1}{2}\chi^2-\frac{1}{2}{\rm ln}\left | C_{\rm ex} \right | + {\rm constant},
\end{split}
\end{equation}
where the inverse covariance $C^{-1}_{\rm ex}$ is given by Eq.~\ref{E:C_inv}. 

We use the fit from individual fields for a consistency check. To provide a best estimate using the combination of all the fields that were observed at once, we also fit to the five CIBER fields using the joint likelihood:
\begin{equation}
{\rm ln}\mathcal{L} = \sum_{i=1}^{N_{\rm field}} {\rm ln}\mathcal{L}_i
\end{equation}
where $N_{\rm field}=5$. Note that the PSF model is different for each field, so the information from different fields is combined in the likelihood.

We use the affine-invariant MCMC sampler \texttt{emcee} \citep{2013PASP..125..306F} to sample from
the posterior distribution. We set flat priors for $R_{\rm e,2}$, $A_{\rm 1h}$, and $A_{\rm 2h}$ in the range of [10$^{-4}R_{\rm 200}$, $R_{\rm 200}$], [0, 50], and [0, 200], respectively. We use an ensemble of 100 walkers taking 1000 steps with 150 burn-in steps. We checked that the chains show good convergence by computing the Gelman-Rubin statistic R \citep{1992StaSc...7..457G}. For all three parameters in all cases, we find $R<1.1$.

\section{Results}\label{S:Results}
\begin{deluxetable*}{c|ccc|ccccc}[th!]
\tablenum{3}
\label{T:param_fit}
\tablecaption{Summary of Parameter Constraints from the Joint Fit in Each Case Listed in Table~\ref{T:mag_bins}}
\tablewidth{0pt}
\tablehead{
\colhead{} &  \colhead{1.1 $\mu$m} & \colhead{1.1 $\mu$m} & \colhead{1.1 $\mu$m} &  \colhead{1.8 $\mu$m} & \colhead{1.8 $\mu$m} & \colhead{1.8 $\mu$m}
\vspace{-0.5em}\\
\colhead{Name} & \colhead{$R_{e,2}$ [arcsec]} & \colhead{$A_{\rm 1h}$} & \colhead{$A_{\rm 2h}$} &  \colhead{$R_{e,2}$ [arcsec]} & \colhead{$A_{\rm 1h}$} & \colhead{$A_{\rm 2h}$}}
\startdata
mag bin \#1 & $<2.76$ & $<6.06$ & $<48.91$
& $<2.53$ & $<5.72$ & $<58.05$\\
mag bin \#2 & 2.25$^{+0.14}_{-0.23}$ & $<4.70$ & $<24.22$
& 1.94$^{+0.12}_{-0.16}$ & $<3.44$ & $<24.76$\\
mag bin \#3 & 1.85$^{+0.17}_{-0.28}$ & $<4.18$ & $<18.94$
& 1.94$^{+0.16}_{-0.16}$ & $<2.96$ & $<18.30$\\
mag bin \#4 & 1.85$^{+0.25}_{-0.21}$ & $<1.16$ & $<6.87$
& 1.63$^{+0.21}_{-0.14}$ & 0.77$^{+0.23}_{-0.23}$ & $<6.59$\\ \hline
total & 1.98$^{+0.17}_{-0.17}$ & $<1.41$* & $<7.30$
& 1.85$^{+0.08}_{-0.15}$ & 1.01$^{+0.24}_{-0.24}$ & $<6.86$\\ \hline
high-M/low-z & 2.30$^{+0.16}_{-0.29}$ & $<4.76$ & $<25.58$
& 2.17$^{+0.18}_{-0.18}$ & $<4.2$ & $<33.10$\\
high-M/med-z & 2.27$^{+0.37}_{-0.32}$ & $<6.42$ & $<19.53$
& 2.22$^{+0.19}_{-0.28}$ & 3.37$^{+1.99}_{-1.17}$ & $<22.76$\\
high-M/high-z & 1.98$^{+0.30}_{-0.44}$ & $<1.88$ & $<9.08$
& 1.85$^{+0.26}_{-0.22}$ & 1.39$^{+0.43}_{-0.35}$ & $<6.19$\\
low-M/low-z & 1.98$^{+0.18}_{-0.30}$ & $<3.18$ & $<16.38$
& 1.89$^{+0.21}_{-0.17}$ & $<2.77$ & $<17.65$\\
low-M/med-z & 1.67$^{+0.29}_{-0.36}$ & $<1.30$ & $<11.30$
& 1.50$^{+0.21}_{-0.24}$ & $<1.01$ & $<7.58$
\enddata
\tablecomments{For the cases with less than a 2$\sigma$ detection (95\% confidence interval), we quote the 2$\sigma$ upper bound. For detections, the $+/-$ values enclose the 68\% confidence interval. In 1.1 $\mu$m ``total'' bin, the 68\% confidence interval of one-halo amplitude $A_{\rm 1h}$ is 0.54$^{+0.42}_{-0.38}$, approximately an $1\sigma$ detection.}
\end{deluxetable*}

\begin{figure*}[ht!]
\includegraphics[width=\linewidth]{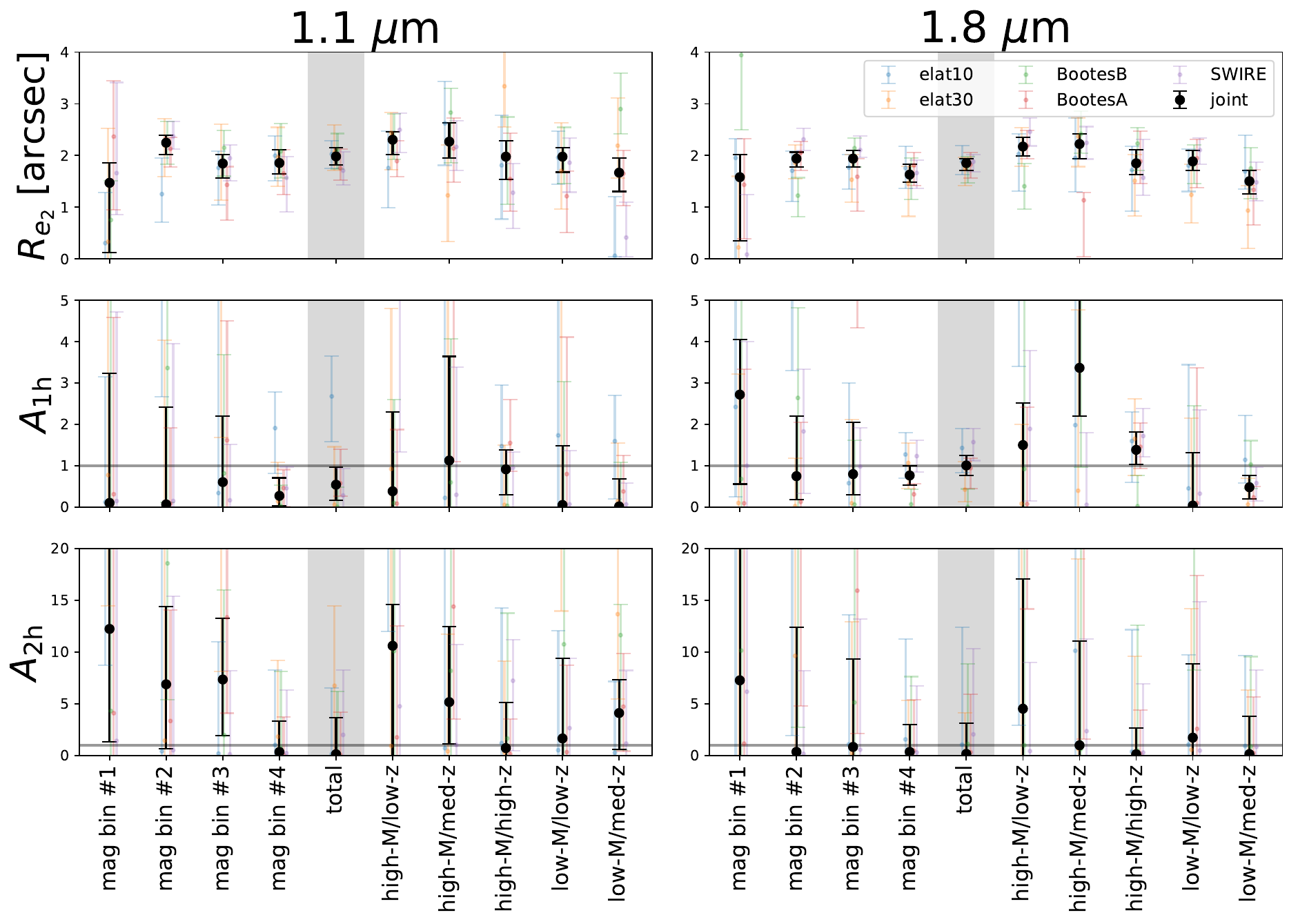}
\caption{\label{F:params_fit} Marginalized parameter constraints from MCMC for each case listed in Table~\ref{T:mag_bins}. The data points and error bars are the median and 68\% confidence intervals from MCMC. Black data points show the joint fit from all five fields, with colored points for the individual fields. The gray horizontal lines in the middle and bottom panels mark $A_{\rm 1h}=1$ and $A_{\rm 2h}=1$, which are the clustering amplitudes given by MICECAT. The shaded regions show the total stack over all $17<m_{\rm 1.1}<20$ galaxies.}
\end{figure*}

\begin{figure}[ht!]
\includegraphics[width=\linewidth]{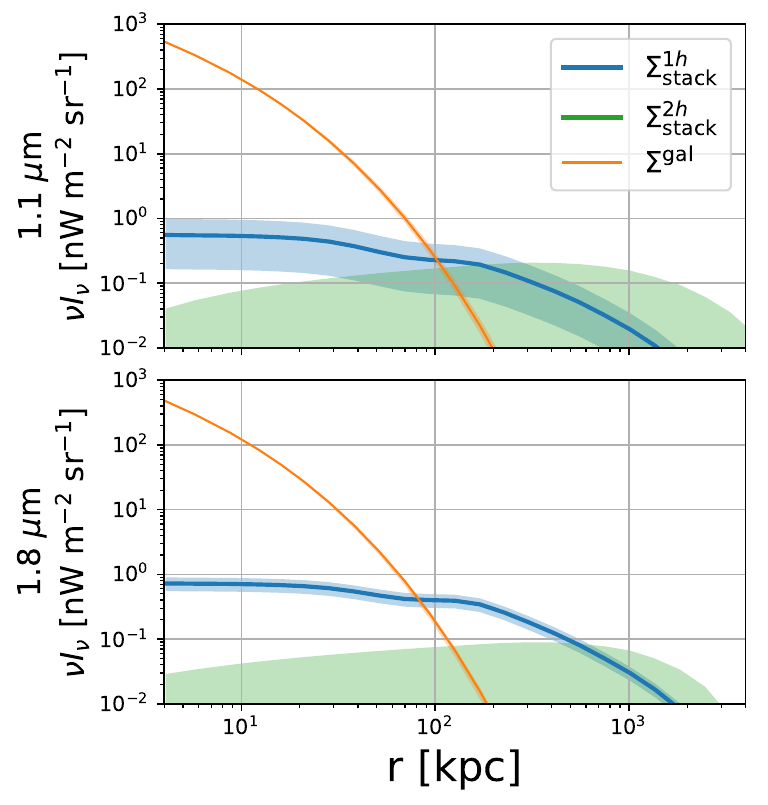}
\caption{\label{F:gal_profin_fit} Fitted intrinsic galaxy profile $\Sigma^{\rm gal}$ (Eq.~\ref{E:Sersic}) (orange), stacked one-halo (blue) and two-halo (green) profiles in the ``total'' magnitude bin averaged over five CIBER fields in the 1.1 $\mu$m (top) and 1.8 $\mu$m (bottom) bands. We convert the angular scale to physical units (kpc) using the median conversion factor inferred from MICECAT (Table~\ref{T:mag_bins}). Solid lines and shaded regions indicate the median and 68\% confidence interval of the joint fit constrained through MCMC, respectively.}
\end{figure}

We show the MCMC results in Fig.~\ref{F:params_fit} and Table~\ref{T:param_fit}, for all cases listed in Table~\ref{T:mag_bins}. As a sanity check, we calculate the $\chi^2$ value between the results from individual fields and the joint fit using 100 data points for each of the three parameters (5 fields $\times$ 10 mag bins $\times$ 2 bands). The resulting $\chi^2$ values indicate our fit is internally consistent across the 5 CIBER fields. In Fig.~\ref{F:stack_profile_fit} and Fig.~\ref{F:excess_profile_fit}, we show the stacked and excess profile data averaged over five fields, respectively, along with the marginalized one-halo, two-halo, and galaxy profile model from the joint fit. Fig.~\ref{F:gal_profin_fit} shows the fitted intrinsic galaxy profile $\Sigma^{\rm gal}$ (Eq.~\ref{E:Sersic}) and the one- and two- halo terms in the ``total'' magnitude bin, also averaged over five fields. The field-averaged profiles are only shown for visualization purposes; when we fit the data with MCMC, the information is combined in the likelihood function rather than in data space.

\section{Discussion}\label{S:Discussion}

\subsection{Missing Light in Galaxy Photometry}
Given the best-fitting extended galaxy profile, we can calculate the fraction of flux missed in photometric galaxy surveys using a limited aperture. From our model, the fraction of flux within a photometric aperture can be approximated by $f_{\rm core} \equiv L_{\rm core}/(L_{\rm core}+L_{\rm ext})$, where $L_{\rm core}$ and $L_{\rm ext}$ are the total flux in the core and extension profile (Eq.~\ref{E:Sigma_core_ext}), respectively. In practice, there are various ways to perform photometry. The Petrosian flux \citep{1976ApJ...209L...1P} is derived from aperture photometry and thus it is the most straightforward method to compare to our results. The Petrosian flux is defined by the total flux within a multiplicative factor of the Petrosian radius of sources. We obtain the Petrosian radius and Petrosian flux from the SDSS catalog of each stacked galaxy in our sample. In SDSS, the Petrosian flux is calculated by integrating the emission within twice the Petrosian radius\footnote{\url{https://www.sdss.org/dr12/algorithms/magnitudes/\#mag_petro}}. With our galaxy profile, we can calculate the fraction of flux within the same radius ($f_{\rm petro}$). The results are summarized in Table~\ref{T:f_SDSS}. 

We also estimate the missing light fraction with the ``model magnitude'' given in SDSS ($f_{\rm model}$). Rather than integrating within a certain aperture size, the model magnitude is derived by fitting the galaxy profile with an exponential or de Vaucouleurs functional form, choosing the one with the higher likelihood in the fitting\footnote{See \url{https://www.sdss.org/dr12/algorithms/magnitudes/} for the detailed descriptions of model magnitudes.}. While it is difficult to apply the same fitting procedure to the sources in CIBER images, we can calculate the ratio between the model flux and the Petrosian flux of each source in the SDSS catalog, and thus infer the fraction of missing light in the model flux. We find that both the Petrosian flux, which measures source emission within a limited aperture size, and the model flux derived from fitting a light profile to the small-radii regions of the galaxy, miss $\sim$ 20\% of the total galaxy light, a deficit detected at $\sim 7\sigma$ ($\sim 4\sigma$) level for Petrosian (model) flux when combing constraints from all five sub-samples. This value is slightly larger than the light fraction in our galaxy extension term ($\sim$ 10 to 20 \%). Our results on the missing light fraction in the Petrosian flux are in agreement with previous analytical calculation \citep{2005AJ....130.1535G}. Interestingly, \citet{2011ApJ...731...89T} probed the radial profile of $z\sim 0.34$ luminous red galaxies (LRGs) in SDSS with a stacking analysis, and they also found $\sim$20\% of the total light missing at large radii when fitting a Sersic model to individual galaxies. Although their galaxy samples are at somewhat higher mass ($M_*\sim 10^{11}-10^{12}M_\odot$) and model magnitudes are fitted with a different functional form, we arrive at a similar fraction of missing flux.

\begin{deluxetable*}{c|ccc|ccc}[ht!]
\tablenum{4}
\label{T:f_SDSS}
\tablecaption{Fraction of flux in core component compared to flux captured in Petrosian and SDSS model flux, assuming the galaxy light profile follows the stacking results in this work. The total row shows the weighted average of the five listed sub-samples.}
\tablewidth{0pt}
\tablehead{
\colhead{} &  \colhead{1.1 $\mu$m} & \colhead{1.1 $\mu$m} & \colhead{1.1 $\mu$m} &  \colhead{1.8 $\mu$m} &  \colhead{1.8 $\mu$m} &  \colhead{1.8 $\mu$m}
\vspace{-0.5em}\\
\colhead{Name} & \colhead{$f_{\rm core}$} & \colhead{$f_{\rm petro}$}  & \colhead{$f_{\rm model}$} &  \colhead{$f_{\rm core}$} & \colhead{$f_{\rm petro}$}  & \colhead{$f_{\rm model}$} 
}
\startdata
high-M/low-z & 0.79$^{+0.04}_{-0.02}$ & 0.78$^{+0.08}_{-0.10}$ & 0.84$^{+0.11}_{-0.12}$ & 0.81$^{+0.02}_{-0.02}$ & 0.80$^{+0.07}_{-0.10}$ & 0.85$^{+0.10}_{-0.12}$\\
high-M/med-z & 0.81$^{+0.04}_{-0.05}$ & 0.74$^{+0.07}_{-0.13}$ & 0.78$^{+0.08}_{-0.15}$ & 0.83$^{+0.04}_{-0.03}$&  0.75$^{+0.08}_{-0.11}$ & 0.78$^{+0.10}_{-0.13}$ \\
high-M/high-z & 0.86$^{+0.06}_{-0.04}$ & 0.73$^{+0.07}_{-0.16}$ & 0.77$^{+0.15}_{-0.19}$ & 0.89$^{+0.03}_{-0.04}$ & 0.75$^{+0.07}_{-0.16}$ & 0.79$^{+0.16}_{-0.18}$\\
low-M/low-z & 0.84$^{+0.04}_{-0.02}$ & 0.78$^{+0.05}_{-0.11}$ & 0.80$^{+0.10}_{-0.12}$ & 0.85$^{+0.02}_{-0.03}$ & 0.79$^{+0.05}_{-0.11}$ & 0.81$^{+0.10}_{-0.12}$\\
low-M/med-z & 0.89$^{+0.04}_{-0.04}$ & 0.78$^{+0.06}_{-0.16}$ & 0.80$^{+0.09}_{-0.16}$ & 0.92$^{+0.03}_{-0.03}$ & 0.80$^{+0.06}_{-0.15}$ & 0.83$^{+0.10}_{-0.14}$\\\hline
total & 0.83$^{+0.02}_{-0.01}$ & 0.77$^{+0.03}_{-0.06}$ & 0.80$^{+0.05}_{-0.06}$ & 0.86$^{+0.01}_{-0.01}$ & 0.78$^{+0.03}_{-0.05}$ & 0.81$^{+0.05}_{-0.06}$\\
\enddata
\end{deluxetable*}

\subsection{Extended Stellar Halo}
\begin{figure*}[ht!]
\includegraphics[width=\linewidth]{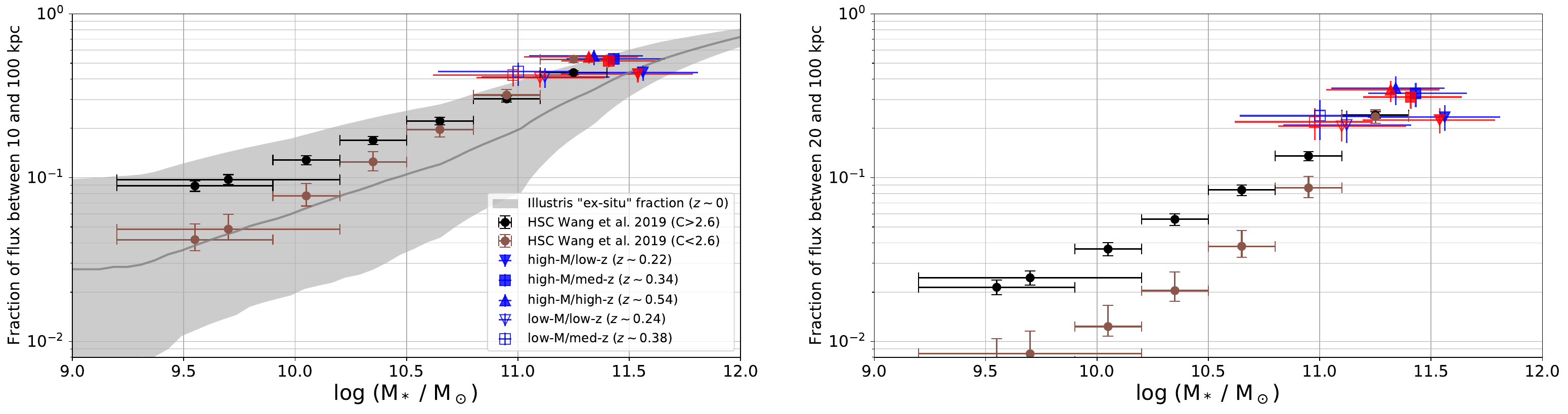}
\caption{\label{F:frac_exsitu} Fraction of flux between 10 (left)/20 (right) and 100 kpc from the galaxy profile derived from CIBER stacking (this work) in the 1.1 (blue) and 1.8 (red) $\mu$m bands and from HSC stacking \citep{2019MNRAS.487.1580W}. The HSC stacking is performed on low and high concentration populations ($C<2.6$ and $C>2.6$) at optical wavelengths (r band). The horizontal error bars define the lower and upper bounds of the stellar mass of each stacking sample. The gray line and the shaded regions in the left panel are the median, 16th, and 84th percentile of the ex situ stellar mass fraction at $z=0$ from Illustris simulations \citep{2016MNRAS.458.2371R}. The shaded region shows the variance between individual galaxies in Illustris, whereas for CIBER and HSC, the error bars represent the standard error on the mean value.}
\end{figure*}

The Illustris simulation \citep{2016MNRAS.458.2371R} traces the dynamics and merger history of stellar particles and estimates the ``ex situ'' population of stars that formed in other galaxies and were later stripped and accreted into a new galaxy. The shaded region in the left panel of Fig.~\ref{F:frac_exsitu} shows the ex situ stellar mass fraction at $z=0$ from the Illustris simulation \citep{2016MNRAS.458.2371R}. Although it is difficult to measure the ex situ component in observations, \citet{2018MNRAS.475.3348H} have studied individual stellar halos out to 100 kpc in more massive galaxies ($10^{11}M_\odot \lesssim M_* \lesssim 10^{12} M_\odot$) at higher redshifts ($z\sim 0.4$) in HSC images, finding that the fraction of stellar mass between 10 and 100 kpc is in good agreement with the ex situ fraction constraints from Illustris \citep{2016MNRAS.458.2371R}. In addition, \citet{2019MNRAS.487.1580W} probe the stellar halo around local ($0\lesssim z\lesssim 0.25$) low-mass galaxies ($9.2\,M_\odot<{\rm log} M_*<11.4\,M_\odot$) with a stacking analysis on HSC images in the \textit{r} band. They stacked galaxies out to $\sim 120$ kpc within several stellar mass bins. For each bin, they split the sources into low and high-concentration populations, defined by $C<2.6$ and $C>2.6$, where $C=R_{\rm 90}/R_{\rm 50}$ is the ratio of the radii that contain 90\% and 50\% of the \textit{r}-band Petrosian flux.

CIBER extends the HSC measurements to higher redshifts and longer wavelength bands. Armed with light profile fits, we can quantify the luminosity fraction in the extended stellar halo around the stacked sources. The left panel of Fig.~\ref{F:frac_exsitu} shows the fraction of stellar flux between radii of 10 and 100 kpc, using the fitted galaxy profile from CIBER and HSC \citep{2019MNRAS.487.1580W}. We observe that $\sim$ 50\% of the flux originates at galactocentric distances between 10 and 100 kpc. \cite{2019MNRAS.487.1580W} re-scaled their images to physical units before stacking, whereas in our analysis we stack sources in observed angular units. Therefore, the variations in our measurements are mostly due to the variation of the conversion factor from angular to physical units for each galaxy in our stack. Our constraints are consistent with the HSC results in the highest mass bin.

Both CIBER and HSC are consistent with the ex situ fraction from Illustris at $z=0$, but are systematically higher than the median value from Illustris (the gray line in Fig.~\ref{F:frac_exsitu}). One possible explanation is that the flux between 10 and 100 kpc is not a perfect proxy of the ex situ population for lower mass galaxies. For example, \citet{2014MNRAS.443.1433D} has shown that the transition scale between in situ and ex situ components varies across a wide range from $\sim 10$ to $\sim 50$ kpc, depending on the stellar mass and concentration of the galaxies. Nevertheless, given the limited information in stacking, we use this definition to associate the luminosity from beyond 10 kpc with IHL.

We also show the fraction of flux with a 20 kpc radius cut in the right panel of Fig.~\ref{F:frac_exsitu}. A radius cut at 20 kpc is a more suitable choice to describe the stripped stellar populations. For example, Milky Way-sized simulations suggest that the infalling stellar debris is recaptured by the galaxy and results in disk thickening at $r\lesssim 10$ kpc \citep{2020A&A...636A.106M}. We note that each galaxy has a different stellar halo profile; interpreting our stacking results requires knowing the stellar halo profile on average over a large sample of galaxies.  Given the uncertainty in choosing an average IHL radius, we report our results in both 10 kpc and 20 kpc scales, while showing the full radial range in Fig.~\ref{F:frac_r}. We find that $\sim$ 25\% of galaxy fluxes are from outside 20 kpc. The CIBER constraints shown in Fig.~\ref{F:frac_exsitu} are summarized in Table~\ref{T:f_ext}.

\subsection{EBL from Extended Stellar Halos}
\begin{figure}[ht!]
\includegraphics[width=\linewidth]{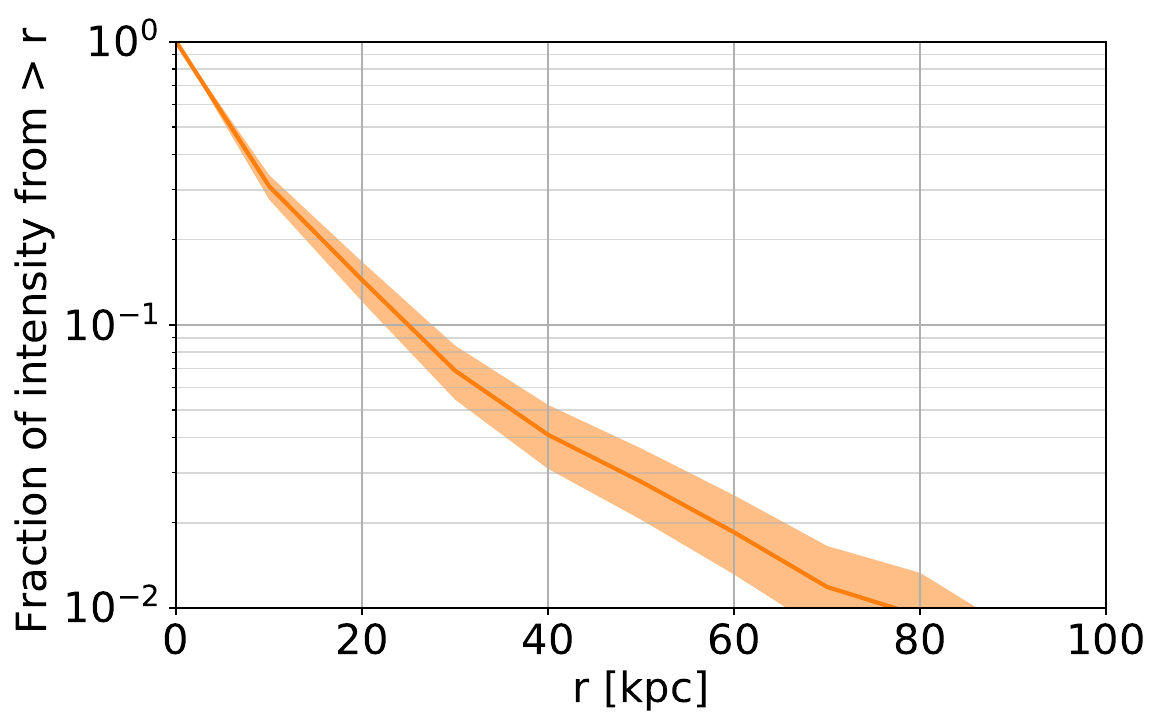}
\caption{\label{F:frac_r} Fraction of EBL intensity from galaxy extension as a function of $r_{\rm cut}$. This is estimated with the light profile fits from CIBER (this work) and HSC \citep{2019MNRAS.487.1580W}, and the stellar mass function from \citet{2013ApJ...777...18M}.}
\end{figure}

With the galaxy profile from CIBER and HSC, we can estimate the EBL contribution from the extended regions at the redshift of our stacked sources. We model this quantity in the following steps: 
\begin{enumerate}
    \item For any given radius cut $r_{\rm cut}$, we model the fraction of light beyond $r_{\rm cut}$ as a function of stellar mass by fitting a line to all CIBER and HSC data points in  logarithmic space;
    \item We estimate total stellar mass density by integrating the stellar mass function from \citet[][]{2013ApJ...777...18M} (we take their single Schechter function fit with all samples in $0.2\leqslant z < 0.5$ bin, approximately the redshift of our sources);
    \item For each $r_{\rm cut}$, we apply the fraction derived in step 1 to the stellar mass function, and integrate to get the stellar mass density from sources outside $r_{\rm cut}$;
    \item Assuming the mass-to-light ratio is the same for all galaxies, the ratio between step 3 and step 2 is our estimate of the EBL fraction from extended sources as a function of $r_{\rm cut}$.
\end{enumerate}
The results are shown in Fig.~\ref{F:frac_r}. We get approximately 30/15 \% of extended emission in the EBL with $r_{\rm cut}=$ 10/20 kpc, respectively. Note that these values are close to the fraction in the five individual stellar mass bins from our stacking results. This is expected as our samples are at $\sim L_*$ scale, which are the representative population that contains the majority of the total stellar emission of their redshift.

\subsection{Intra-halo Light Fraction}\label{S:fIHL}
The fraction of the total emission from a dark matter halo associated with IHL, $f_{\rm IHL}$, has been investigated with both observation and theoretical modeling \citep[e.g., ][]{2004ApJ...617..879L,2005ApJ...618..195G,2007ApJ...666...20P,2014MNRAS.443.1433D,2015MNRAS.449.2353B,2018MNRAS.479.4004E}. With our stacking results, we can estimate the total halo emission from the sum of the galaxy light and one-halo terms. For the IHL, we consider the extended galaxy emission beyond $r_{\rm cut}=$ 10/20 kpc of all the bright ($m_{1.1}<20$) galaxies in the halo, noting that $m_{1.1}=20$ is also our choice of flux threshold for masking. Therefore, the IHL fraction $f_{\rm IHL}$ can be expressed as
\begin{equation}
f_{\rm IHL}=\frac{\sum_{m_{1.1}<20}L (>r_{\rm cut})}{\sum_{m_{1.1}<20}L + \sum_{\rm faint}L},
\end{equation}
where $\sum_{m_{1.1}<20}L$ is the total light associated with bright galaxies, and $\sum_{m_{1.1}<20}L (>r_{\rm cut}$) is the part of bright galaxy emission beyond $r_{\rm cut}$. $\sum_{\rm faint}L$ represents the light from faint galaxies as well as the unbound stars in the halo, captured in the one-halo luminosity.  Note that we conservatively assume the one-halo luminosity arises entirely from faint, gravitationally bound galaxies. However, it is certainly true that some one-halo light arises from unbound stars as is readily observed in images of massive clusters at low redshift.

From our stacking profile, the faint source emission $\sum_{\rm faint}L$ can be described by the total emission in the one-halo term, $L_{\rm 1h}$\footnote{Our one-halo model also includes the outskirts of bright sources beyond the mask, but we checked that this component is negligible compared to the faint sources using the MICECAT simulation.}. For the bright sources, we define
\begin{equation}
\sum_{m_{1.1}<20}L = L_{\rm gal} \cdot N_{\rm eff},
\end{equation}
where $L_{\rm gal}$ is the total light in the galaxy profile term from our stacking results, which describes the averaged light of the galaxies within each stacking sample. $N_{\rm eff}$ accounts for the fact that there are multiple bright galaxies in the halo, and we infer the average $N_{\rm eff}$ value from MICECAT. For our five stacking sub-samples, we get $N_{\rm eff}\sim$ 2--5. From our fitted galaxy profile, we can also calculate $L_{\rm gal} (>r_{\rm cut})$, and we apply the same $N_{\rm eff}$ to model the extension from other bright galaxies:

\begin{equation}
\sum_{m_{1.1}<20}L (>r_{\rm cut})= L_{\rm gal} (>r_{\rm cut}) \cdot N_{\rm eff}.
\end{equation}
This results in
\begin{equation}\label{E:f_IHL}
f_{\rm IHL}=\frac{L_{\rm gal}(>r_{\rm cut})/L_{\rm gal}}{1+L_{\rm 1h}/\left (N_{\rm eff}\cdot L_{\rm gal}  \right )}.
\end{equation}

\begin{figure*}[ht!]
\includegraphics[width=\linewidth]{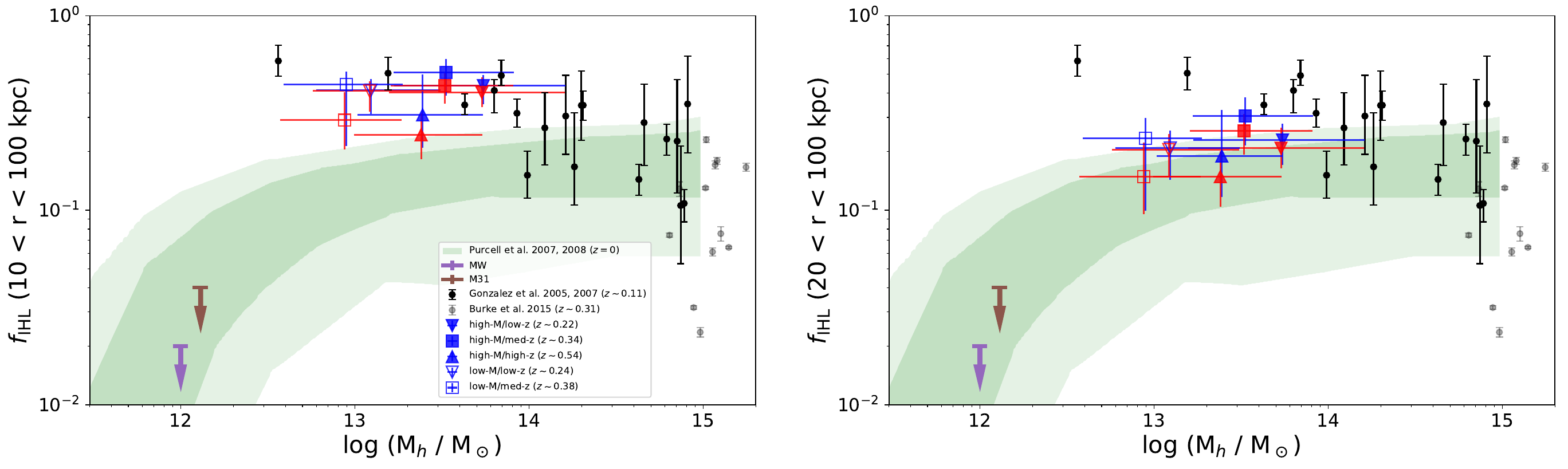}
\caption{\label{F:fIHL}The IHL fraction $f_{\rm IHL}$ as a function of halo mass. The IHL is defined by the light beyond a radius $r_{\rm cut}$ around the galaxy. Here we consider three different $r_{\rm cut}$ values: 10 kpc (left) and 20 kpc (right). Blue and red data points show the constraints from this work in the 1.1 $\mu$m and 1.8 $\mu$m bands, respectively. Dark and light green shaded regions denote the 68\% and 95\% variations among galaxies from an analytical model at $z=0$ \citep{2007ApJ...666...20P, 2008MNRAS.391..550P}. The ICL fraction in individual galaxy groups and clusters from \citet{2005ApJ...618..195G, 2007ApJ...666..147G} and \citet{2015MNRAS.449.2353B} are shown in black and gray data points. The two downward arrows give upper limits for the Milky Way \citep{2010ApJ...712..692C} and Andromeda (M31) \citep{2011ApJ...739...20C}.}
\end{figure*}

\begin{figure*}[ht!]
\includegraphics[width=\linewidth]{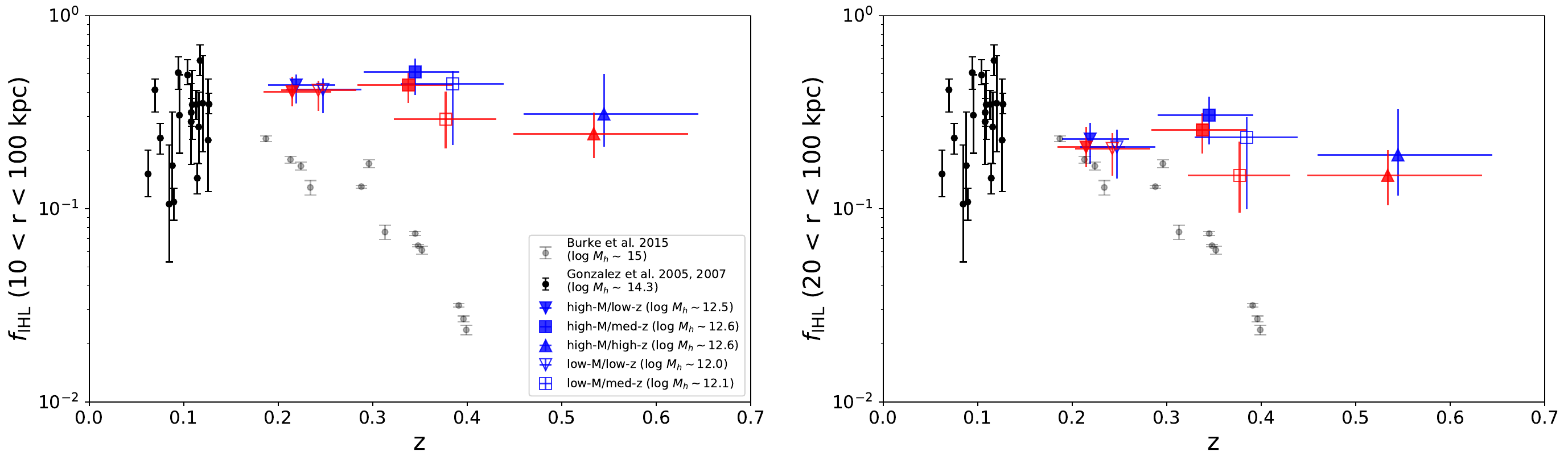}
\caption{\label{F:fIHL_z} $f_{\rm IHL}$ constraints as in Fig.~\ref{F:fIHL}, but plotted as a function of redshift. The masses of the \citet{2015MNRAS.449.2353B} clusters are 100-1000$\times$ the halo masses associated with our galaxies.}
\end{figure*}

We show our constraints on $f_{\rm IHL}$, as a function of halo mass and redshift in Fig.~\ref{F:fIHL} and \ref{F:fIHL_z}, respectively. The halo masses associated with our galaxies are inferred from the MICECAT simulation and using the SDSS photometric redshifts. The CIBER data points shown in Fig.~\ref{F:fIHL} and \ref{F:fIHL_z} are summarized in Table~\ref{T:f_IHL}.

\begin{figure}[ht!]
\includegraphics[width=\linewidth]{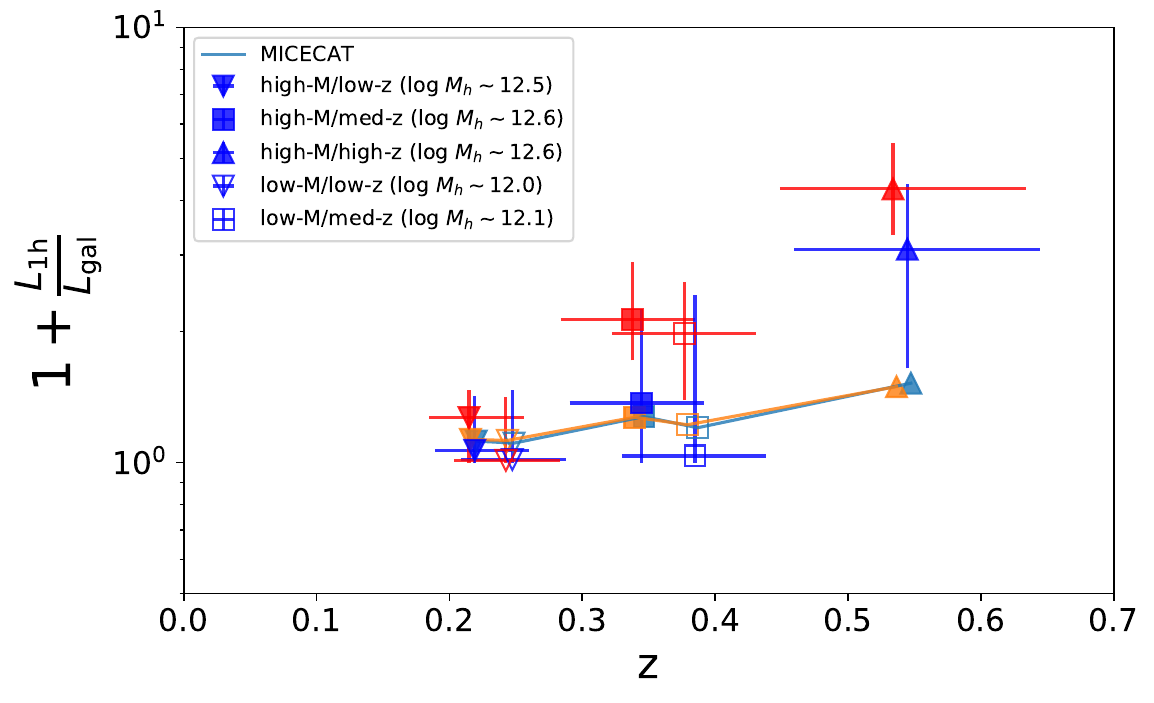}
\caption{\label{F:f1h_gal} Ratio of the total one-halo term and stacked galaxy profile term from our stacking results (blue: 1.1 $\mu$m, red: 1.8 $\mu$m) compared with the MICECAT simulation (light blue: 1.1 $\mu$m, orange: 1.8 $\mu$m). We observe a somewhat stronger evolution, causing the fall-off of $f_{\rm IHL}$ with redshift seen in Fig.~\ref{F:fIHL_z}.}
\end{figure}

Note that the fraction of light beyond $r_{\rm cut}$ (the numerator in Eq.~\ref{E:f_IHL}) is shown in Fig.~\ref{F:frac_exsitu}, where the higher redshift bins have slightly higher values. However, in Fig.~\ref{F:fIHL_z}, they have lower $f_{\rm IHL}$. This is due to the increase of the one-halo term with redshift. We show the ratio of the one-halo term and the stacked galaxy light in Fig.~\ref{F:f1h_gal}. Note that this observable quantity tracks the evolution of the one-halo luminosity but lacks the $N_{\rm eff}$ term in Eq.~\ref{E:f_IHL} derived from simulations. We compare with the same quantity from the MICECAT simulation, where the one-halo term includes all the unmasked faint galaxies and residual bright source emission outside the mask due to the PSF. We detect a strong redshift evolution of one-halo contribution compared with the MICECAT simulation, which could be attributed to the unbound stars that are not included in MICECAT.

We compare our results with $f_{\rm IHL}$ from previous work, including the Milky Way \citep{2010ApJ...712..692C}, the Andromeda galaxy \citep[M31; ][]{2011ApJ...739...20C}, the ICL fraction in individual galaxy groups and clusters \citep{2005ApJ...618..195G, 2007ApJ...666..147G,2015MNRAS.449.2353B}, and an analytical model \citep{2007ApJ...666...20P, 2008MNRAS.391..550P}. Our results follow a more gradual redshift evolution trend than reported in massive clusters \citep[][see Fig.~\ref{F:fIHL_z}]{2015MNRAS.449.2353B}.

\subsection{Color of the Galaxy Inner and Outer Regions}
We calculate the $m_{\rm 1.1}-m_{\rm 1.8}$ color of the inner and outer region of the galaxy, defined by the total light inside and outside 20 kpc physical scale in the fitted galaxy profile. The results are summarized in Table~\ref{T:color}. Note the definition of inner and outer component here is based on the intrinsic profile, which is different from the core/extension separation using the stacked PSF defined in Eq.~\ref{E:Sigma_core_ext}. We have no detection of a color difference between the inner and outer regions in the two CIBER bands. We also find similar inner and outer region color with 10 kpc radius cut. Previous measurements in optical bands found that the galaxy outskirts are bluer than their core \citep[e.g., ][]{2014MNRAS.443.1433D,2018MNRAS.475.3348H}. For comparison, we calculate the $m_{\rm 1.1}-m_{\rm 1.8}$ color of galaxy cores in MICECAT sources selected from the same criteria, as well as from the empirical galaxy model of \citet{2012ApJ...752..113H} at $z=0.3$, approximately the redshift of our samples. Our inner region color is consistent with these models. To model the extension, we use a collection of elliptical galaxy spectra from the population synthesis package GISSEL \citep{1993ApJ...405..538B} redshifted to $z=0.3$. We also estimate the extension color using an imaging study on the local spiral galaxy NGC 5907 \citep{1997Natur.387..159R}. We use their ratio of $I$ band and $J$ band flux in $>$1 arcmin regions to approximate the $m_{\rm 1.1}-m_{\rm 1.8}$ extension color. The rest-frame $I$ and $J$ band redshifted to $z\sim 0.3$ (approximately the redshift of our samples) are close to the two CIBER bands. NGC 5907 shows a redder spectrum than our galaxy extension, whereas the elliptical galaxy spectrum template is slightly bluer than our samples. In addition, the IHL constraints from \citet{2014Sci...346..732Z} are also given in Table~\ref{T:color}, but we note that \citet{2014Sci...346..732Z} reflects the integrated IHL from all redshifts.

\begin{deluxetable}{c|cc}[h]
\tablenum{5}
\label{T:color}
\tablecaption{Constraints on the Color ($m_{\rm 1.1}-m_{\rm 1.8}$) of the Galaxy Inner and Outer Components}
\tablewidth{0pt}
\tablehead{
\colhead{Name} & \colhead{Inner} & \colhead{Outer}}
\startdata
high-M/low-z & 0.42$^{+0.20}_{-0.17}$ & 0.36$^{+0.34}_{-0.31}$\\
high-M/med-z & 0.54$^{+0.25}_{-0.27}$ & 0.46$^{+0.24}_{-0.25}$\\
high-M/high-z & 0.65$^{+0.31}_{-0.28}$ & 0.61$^{+0.31}_{-0.25}$\\
low-M/low-z & 0.39$^{+0.20}_{-0.18}$ & 0.37$^{+0.41}_{-0.37}$\\
low-M/med-z & 0.56$^{+0.23}_{-0.24}$ & 0.44$^{+0.50}_{-0.44}$\\ \hline
total & 0.49$^{+0.10}_{-0.10}$ & 0.47$^{+0.14}_{-0.15}$\\\hline\hline
MICECAT & 0.44$\pm$0.07\\
Helgason et al. (2012)  & 0.41 \\
GISSEL & & 0.32$\pm$ 0.08  \\
NGC 5907 & & 1.41$\pm$ 0.61  \\\hline
Zemcov et al. (2014) & & 0.89$^{+1.17}_{-1.08}$
\enddata
\tablecomments{The $+/-$ values indicate 68\% interval ranges. The total row shows the weighted average of five sub-samples. For comparison, we also show models of core color from MICECAT and an analytical prescription from \citet{2012ApJ...752..113H} at $z=0.3$. For the extension, we compare our results with spectra from a population synthesis code, GISSEL \citep{1993ApJ...405..538B}, and the outskirts of NGC 5907 redshifted to $z=0.3$ \citep{1997Natur.387..159R}. The color of EBL fluctuations attributed to redshift-integrated IHL from \citet{2014Sci...346..732Z} is also shown.}
\end{deluxetable}

\subsection{One-halo and Two-halo Clustering}
The one-halo amplitude is detected in the 1.8 $\mu$m band at the $\sim 4\sigma$ level in the ``total'' and ``high-M/high-z'' cases, and at the $\sim 3\sigma$ level in ``mag bin \#4'' and ``high-M/med-z'' cases. One-halo clustering is not clearly detected at the 1.1 $\mu$m band since the photocurrent from sources is lower in this band. The one-halo amplitude $A_{\rm 1h}$ is consistent with unity to within $\sim 2 \sigma$, which implies that our one-halo templates built from MICECAT are sufficient to describe the clustering within halos of our stacked samples. However, from our stacking results, it is unclear if this emission actually consists of discrete galaxies as given in the MICECAT simulation. Two-halo clustering is not detected in all cases since the large-scale clustering signal is comparable to the current uncertainties in the measurement.

\section{Conclusions}\label{S:Conclusion}
By stacking galaxies from CIBER imaging data in two near-infrared bands (1.1 and 1.8 $\mu$m), we detect extended emission in galaxies. The galaxies being stacked ($\sim 30,000$ galaxies in total) are split into five sub-samples from SDSS spanning redshifts $0.2 \lesssim z \lesssim 0.5$ and stellar masses $10^{10.5}M_\odot \lesssim M_* \lesssim 10^{12} M_\odot$, comparable to $L_*$ galaxies at this redshift. We jointly fit a model for the inherent galaxy light profile and large-scale one- and two-halo clustering.

With the galaxy profile, we estimate that $\sim 20\%$ of total light is missing in galaxy photometry due to the use of limited apertures, in agreement with previous estimates from the literature. We do not detect a 1.1-1.8 $\mu$m color difference in the inner and outer region of our galaxy samples.

While we do not detect two-halo clustering, we detect one-halo clustering in the 1.8 $\mu$m band at 4$\sigma$ significance over the full sample of galaxies. These results suggest nonlinear clustering could have a significant impact on modeling the IHL, but is not accounted for in previous fluctuation analysis by \citet{2014Sci...346..732Z}. An IHL fluctuation model with one-halo clustering \citep[e.g.,][]{2010ApJ...710.1089F} is needed to fully account for the nonlinear clustering in IHL modeling.

The intrinsic galaxy profile fitted from our stacking analysis suggests $\sim50\%$/$25\%$ of the total galaxy light resides at $r>$ 10/20 kpc, respectively. This result is in agreement with previous HSC measurements at lower redshifts ($0\lesssim z \lesssim 0.25$) and lower stellar masses ($10^{9.2}M_\odot < M_* < 10^{11.4} M_\odot$). The galaxy extension accounts for significant fraction of luminosity in $L_*$ galaxies, but falls off below $M_*\sim 10^{11}M_\odot$. We extrapolate the fraction of extended galaxy light we measure to all galaxy mass scales and assuming a Schechter luminosity function, we find $\sim$ 30\%/15\% of the total galaxy light are from $r >$ 10/20 kpc, respectively. We measure a moderate increase in $f_{\rm IHL}$ with cosmic time, which we attribute to the decrease in one-halo contribution within the dark matter halo of our stacked samples. The previous fluctuation study using CIBER data \citep{2014Sci...346..732Z} found that the IHL has comparable intensity to the IGL in the near-infrared EBL. While our study cannot constrain the whole IHL contribution to the EBL since we only study galaxies from a certain range of redshift and masses, our results suggest that $\sim L_*$ galaxy at $0.2\lesssim z \lesssim 0.5$ have an extended light profile which contributes appreciable IHL to their host halos. As $\sim L_*$ galaxies are the representative population, which contain most of the IGL emission, the flux from the extension, and the one-halo term present in our galaxy samples, both need to be properly accounted for in future EBL photometry and fluctuation measurements.

\acknowledgments
We would like to thank the anonymous referee for valuable comments that improved the manuscript. We thank the dedicated efforts of the sounding rocket staff at NASA Wallops Flight Facility and White Sands Missile Range. This work was supported by NASA APRA research grants NNX07AI54G, NNG05WC18G, NNX07AG43G, NNX07AJ24G, NNX10AE12G, and 80NSSC20K0595. Initial support was provided by an award to J.B. from the Jet Propulsion Laboratory’s Director’s Research and Development Fund. Japanese participation in CIBER was supported by KAKENHI (20.34, 18204018, 19540250, 21340047, 21111004, 24111717, 26800112, and 15H05744) from Japan Society for the Promotion of Science (JSPS) and the Ministry of Education, Culture, Sports, Science and Technology (MEXT). Korean participation in CIBER was supported by the Pioneer Project from Korea Astronomy and Space Science Institute (KASI). Y.-T.C. acknowledges support by the Ministry of Education, Taiwan through the Taiwan-Caltech Scholarship. C.H.N. acknowledges support by NASA Headquarters under the NASA Earth and Space Science Fellowship Program - Grant 80NSSCK0706. Part of the research was carried out at the Jet Propulsion Laboratory, California Institute of Technology, under a contract with the National Aeronautics and Space Administration (80NM0018D0004).

This publication makes use of data products from the Two Micron All Sky Survey, which is a joint project of the University of Massachusetts and the Infrared Processing and Analysis Center/California Institute of Technology, funded by the National Aeronautics and Space Administration and the National Science Foundation.

The Pan-STARRS1 Surveys (PS1) and the PS1 public science archive have been made possible through contributions by the Institute for Astronomy, the University of Hawaii, the Pan-STARRS Project Office, the Max-Planck Society and its participating institutes, the Max Planck Institute for Astronomy, Heidelberg and the Max Planck Institute for Extraterrestrial Physics, Garching, The Johns Hopkins University, Durham University, the University of Edinburgh, the Queen's University Belfast, the Harvard-Smithsonian Center for Astrophysics, the Las Cumbres Observatory Global Telescope Network Incorporated, the National Central University of Taiwan, the Space Telescope Science Institute, the National Aeronautics and Space Administration under Grant No. NNX08AR22G issued through the Planetary Science Division of the NASA Science Mission Directorate, the National Science Foundation grant No. AST-1238877, the University of Maryland, Eotvos Lorand University (ELTE), the Los Alamos National Laboratory, and the Gordon and Betty Moore Foundation.

Funding for the Sloan Digital Sky Survey IV has been provided by the Alfred P. Sloan Foundation, the U.S. Department of Energy Office of Science, and the Participating Institutions. SDSS-IV acknowledges
support and resources from the Center for High-Performance Computing at
the University of Utah. The SDSS website is www.sdss.org.

SDSS-IV is managed by the Astrophysical Research Consortium for the 
Participating Institutions of the SDSS Collaboration including the 
Brazilian Participation Group, the Carnegie Institution for Science, 
Carnegie Mellon University, the Chilean Participation Group, the French Participation Group, Harvard-Smithsonian Center for Astrophysics, 
Instituto de Astrof\'isica de Canarias, The Johns Hopkins University, Kavli Institute for the Physics and Mathematics of the Universe (IPMU) / 
University of Tokyo, the Korean Participation Group, Lawrence Berkeley National Laboratory, 
Leibniz Institut f\"ur Astrophysik Potsdam (AIP),  
Max-Planck-Institut f\"ur Astronomie (MPIA Heidelberg), 
Max-Planck-Institut f\"ur Astrophysik (MPA Garching), 
Max-Planck-Institut f\"ur Extraterrestrische Physik (MPE), 
National Astronomical Observatories of China, New Mexico State University, 
New York University, University of Notre Dame, 
Observat\'ario Nacional / MCTI, The Ohio State University, 
Pennsylvania State University, Shanghai Astronomical Observatory, 
United Kingdom Participation Group,
Universidad Nacional Aut\'onoma de M\'exico, University of Arizona, 
University of Colorado Boulder, University of Oxford, University of Portsmouth, 
University of Utah, University of Virginia, University of Washington, University of Wisconsin, 
Vanderbilt University, and Yale University.

This work has made use of data from the European Space Agency (ESA) mission
 Gaia (\url{https://www.cosmos.esa.int/gaia}), processed by the Gaia
Data Processing and Analysis Consortium (DPAC,
\url{https://www.cosmos.esa.int/web/gaia/dpac/consortium}). Funding for the DPAC
has been provided by national institutions, in particular the institutions
participating in the Gaia Multilateral Agreement.

This work has made use of CosmoHub. CosmoHub has been developed by the Port d'Informaci{\'o} Cient{\'i}fica (PIC), maintained through a collaboration of the Institut de F{\'i}sica d'Altes Energies (IFAE) and the Centro de Investigaciones Energ{\'e}ticas, Medioambientales y Tecnol{\'o}gicas (CIEMAT) and the Institute of Space Sciences (CSIC \& IEEC), and was partially funded by the "Plan Estatal de Investigaci{\'o}n Científica y Técnica y de Innovación" program of the Spanish government.

\software{astropy \citep{2013A&A...558A..33A},  
          emcee \citep{2013PASP..125..306F},
          corner \citep{corner}, astrometry.net \citep{2010AJ....139.1782L},  LePHARE \citep{1999MNRAS.310..540A,2006A&A...457..841I}
          }

\appendix

\section{Extension and IHL Fraction}
Table~\ref{T:f_ext} summarizes the fraction of light beyond 10 and 20 kpc, assuming our fitted light profile. These are the data presented in Fig.~\ref{F:frac_exsitu}.

\begin{deluxetable*}{c|cc|cc}[ht!]
\tablenum{6}
\label{T:f_ext}
\tablecaption{Fraction of Galaxy Flux Between 10/20 kpc and 100 kpc, Assuming the Galaxy Light Profile Follows the Stacking Results in This Work}
\tablewidth{0pt}
\tablehead{
\colhead{} &  \colhead{1.1 $\mu$m} & \colhead{1.1 $\mu$m} &  \colhead{1.8 $\mu$m} &  \colhead{1.8 $\mu$m}
\vspace{-0.5em}\\
\colhead{Name} & \colhead{10 kpc} & \colhead{20 kpc} &  \colhead{10 kpc} & \colhead{20 kpc}
}
\startdata
high-M/low-z & 0.44$^{+0.05}_{-0.05}$ & 0.23$^{+0.04}_{-0.04}$ & 0.43$^{+0.05}_{-0.04}$ & 0.22$^{+0.04}_{-0.04}$\\
high-M/med-z & 0.53$^{+0.05}_{-0.04}$ & 0.31$^{+0.05}_{-0.04}$ & 0.52$^{+0.05}_{-0.04}$&  0.30$^{+0.04}_{-0.04}$ \\
high-M/high-z & 0.55$^{+0.07}_{-0.05}$ & 0.33$^{+0.06}_{-0.05}$ & 0.55$^{+0.05}_{-0.04}$ & 0.33$^{+0.05}_{-0.04}$\\
low-M/low-z & 0.41$^{+0.06}_{-0.05}$ & 0.21$^{+0.04}_{-0.04}$ & 0.41$^{+0.05}_{-0.05}$ & 0.20$^{+0.04}_{-0.04}$\\
low-M/med-z & 0.45$^{+0.08}_{-0.06}$ & 0.23$^{+0.06}_{-0.05}$ & 0.42$^{+0.06}_{-0.05}$ & 0.21$^{+0.05}_{-0.04}$\\\hline
total & 0.48$^{+0.02}_{-0.03}$ & 0.25$^{+0.02}_{-0.02}$ & 0.47$^{+0.02}_{-0.02}$ & 0.25$^{+0.02}_{-0.02}$\\
\enddata
\tablecomments{These are the values shown in Fig.~\ref{F:frac_exsitu}.
The total row shows the weighted average of the five listed sub-samples.}
\end{deluxetable*}

Table~\ref{T:f_IHL} summarize the $f_{\rm IHL}$ values with $r_{\rm cut=}$ 10 and 20 kpc, assuming our fitted light profile and the one-halo contribution from the MICECAT. These are the data presented in Fig.~\ref{F:fIHL} and \ref{F:fIHL_z}.

\begin{deluxetable*}{c|cc|cc}[ht!]
\tablenum{7}
\label{T:f_IHL}
\tablecaption{IHL Fraction (Eq.~\ref{E:f_IHL}) with $r_{\rm cut=}$ 10/20 kpc, Assuming the Galaxy Light Profile and the One-halo Terms Follow Our Stacking Results and the MICECAT Simulation, Respectively}
\tablewidth{0pt}
\tablehead{
\colhead{} &  \colhead{1.1 $\mu$m} & \colhead{1.1 $\mu$m} &  \colhead{1.8 $\mu$m} &  \colhead{1.8 $\mu$m}
\vspace{-0.5em}\\
\colhead{Name} & \colhead{10 kpc} & \colhead{20 kpc} &  \colhead{10 kpc} & \colhead{20 kpc}
}
\startdata
high-M/low-z & 0.44$^{+0.09}_{-0.06}$ & 0.23$^{+0.06}_{-0.05}$ & 0.40$^{+0.06}_{-0.08}$ & 0.21$^{+0.04}_{-0.06}$\\
high-M/med-z & 0.51$^{+0.12}_{-0.09}$ & 0.30$^{+0.09}_{-0.08}$ & 0.44$^{+0.08}_{-0.07}$&  0.26$^{+0.06}_{-0.06}$\\
high-M/high-z & 0.31$^{+0.10}_{-0.19}$ & 0.19$^{+0.07}_{-0.14}$ & 0.24$^{+0.06}_{-0.07}$ & 0.15$^{+0.04}_{-0.05}$\\
low-M/low-z & 0.41$^{+0.10}_{-0.06}$ & 0.21$^{+0.07}_{-0.05}$ & 0.41$^{+0.09}_{-0.05}$ & 0.21$^{+0.06}_{-0.04}$\\
low-M/med-z & 0.44$^{+0.23}_{-0.07}$ & 0.23$^{+0.14}_{-0.06}$ & 0.29$^{+0.09}_{-0.12}$ & 0.15$^{+0.05}_{-0.07}$\\\hline
total & 0.43$^{+0.03}_{-0.05}$ & 0.23$^{+0.03}_{-0.03}$ & 0.36$^{+0.03}_{-0.05}$ & 0.19$^{+0.02}_{-0.02}$\\
\enddata
\tablecomments{These are the values shown in Fig.~\ref{F:fIHL} and \ref{F:fIHL_z}. The total row shows the weighted average of the five listed sub-samples.}
\end{deluxetable*}

\bibliography{reference}{}
\bibliographystyle{aasjournal}

\end{document}